\documentclass[pre, twocolumn, showpacs,amsmath,amssymb]{revtex4-1}

\usepackage[T1]{fontenc}                                                                                                                                                                                                   
\usepackage[latin9]{inputenc}
\usepackage{graphicx}
\usepackage{dcolumn}
\usepackage{xcolor}
\usepackage{braket}
\usepackage{commath}
\usepackage{amsmath}
\usepackage{adjustbox}
\usepackage{comment} 
\usepackage{ulem}
\usepackage{tikz}
\usetikzlibrary{shapes}
\newcommand*\circled[1]{\tikz[baseline=(char.base)]{
    \node[shape=ellipse,draw,inner sep=2pt] (char) {#1};}}
\newcommand*\squared[1]{\tikz[baseline=(char.base)]{
    \node[shape=rectangle,draw,inner sep=2pt] (char) {#1};}}

\newcommand{\bs}[1]{{\boldsymbol{#1}}}
\newcommand{\bk}{\bs{k}}
\newcommand{\br}{\bs{r}}

\definecolor{mygreen}{RGB}{20,148,20}

\usepackage{filemod}

\begin{document}

\title{Semiclassical spectral function and density of states in speckle potentials}

\author{Tony Prat$^1$, Nicolas Cherroret$^1$, Dominique Delande$^1$}
\affiliation{$^1$Laboratoire Kastler Brossel, UPMC-Sorbonne Universit\'es, CNRS, ENS-PSL Research University, Coll\`ege de France; 4 Place Jussieu, 75005 Paris, France}

%\date{18/07/2016}

\begin{abstract}
We present a novel analytical method for calculating the spectral function and the density of states in speckle potentials, valid in the semiclassical regime. Our approach relies on stationary phase approximations, allowing us to describe the singular quantum corrections at low energies. We apply it to the calculation of the spectral function and the density of states in one and two-dimensional speckle potentials. By connecting our results with those of previous work valid in the high energy sector, we end up with a consistent description of the whole energy spectrum, in good agreement with numerical simulations.
\end{abstract}


\maketitle

\section{Introduction}

Anderson localization, the absence of wave diffusion due to destructive interference between partial waves multiply scattered by a disordered potential \cite{Anderson1958}, has been observed in a number of experiments involving atomic matter waves quasi-periodically kicked by laser pulses \cite{Chabe08, Manai15} or subjected to one-dimensional 1D \cite{Billy2008} and three-dimensional 3D \cite{Jendrzejewski12, Semeghini15} quenched speckle potentials, as well as ultrasound waves in 3D disordered dielectric media \cite{Hu2008}. 
In cold-atom setups, the control of atom-atom interactions (through, e.g., Feshbach resonances) together with a weak coupling to the environment constitute precious assets for the observation of interference effects in disorder. Furthermore, atom-optics experiments offer the possibility to directly probe localization phenomena \emph{inside} the atomic system, as well as to follow their evolution in the course of time \cite{Modugno10, Shapiro12}. 

If atoms are injected into a disordered potential with an initial momentum $\bk$, they no longer have a well defined energy $\epsilon$ but rather an energy distribution called the spectral function, denoted by $A_\bk(\epsilon)$. The spectral function thus defines a quasi-particle, and generally speaking can provide important physical insights to the complex problem of disorder scattering even without the knowledge of the system's eigenstates \cite{Anderson97}. Even more, it turns out that to achieve a quantitative understanding of cold-atom experiments in speckle potentials and in particular to properly characterize Anderson localization, a good knowledge of the spectral function is crucial. Indeed, when disorder is strong enough the spectral function is broad, which can have important consequences for the global motion of an atomic cloud. For instance, a cloud of atoms that are individually diffusive may exhibit a global sub-diffusive behavior as a result of the superposition of the various energy-dependent atomic diffusion coefficients, thus mimicking the onset of localization \cite{McGehee13, Mueller14}. Furthermore, even if the cloud contains localized atoms, usually a finite part of it remains diffusive and a precise characterization of the spectral function is then required in order to pinpoint the location of the mobility edge \cite{Semeghini15,Mueller16}. Related to the spectral function, the density of states (DoS) in strong speckle potentials is also poorly understood. This question is however essential as the DoS plays a central role in atomic physics, in particular in the discussion of phases of interacting bosons \cite{Lugan07, Aleiner10, Cherroret15}.

Despite its importance, the calculation of the spectral function of speckle potentials in the strong disorder regime has been little addressed, the main difficulty stemming from the inapplicability of weak-disorder approximations in this regime. Recently however, a systematic semiclassical expansion of the spectral function around
the classical solution has been proposed \cite{Trappe15}. Although successful in the large-energy sector, the approach of \cite{Trappe15} fails at capturing the singular quantum corrections at low energies. As far as the DoS is concerned, important progress has been recently accomplished by Falco \textit{et al.} \cite{Falco10}, who used a classical approximation for describing high energies in speckle potentials. Again however, this approach remains inaccurate to capture the low-energy sector. As a matter of fact, the difficulty of treating low energies in speckle potentials lies in the singular nature of quantum corrections in this region of the spectrum. Such singular corrections are absent for Gaussian random potentials \cite{Trappe15} frequently used in condensed-matter physics \cite{John88}. To our knowledge, they have not been described yet. 

In this paper, we calculate the spectral function and the density of states in one (1D) and two-dimensional (2D) speckle potentials, making use of a semiclassical approach based on stationary phase approximations, thereby allowing for a non analytic perturbation expansion in $\hbar$. Our theoretical predictions are in good agreement with exact numerical simulations in the low-energy sector where quantum corrections are singular. By connecting our results with those of \cite{Trappe15}, we eventually end up with a consistent description of the whole energy spectrum.  
Section \ref{defandmethods} is devoted to the definition of the relevant quantities and to a discussion of the results previously obtained in \cite{Trappe15}. Our semiclassical approach is also introduced and discussed. 
In Sec. \ref{sec1D}, we derive important statistical properties of 1D speckles needed to implement our semiclassical theory. Results for the 1D spectral function and DoS are presented in Sec. \ref{Sec_results_1D}. The approach is then extended to the 2D case in Secs. \ref{Sec_2D_statistics} and \ref{Sec_results_2D}. In Sec. \ref{secconcl}, we finally summarize our findings and discuss some open questions.

\section{Definitions and methods}
\label{defandmethods}

\subsection{Framework}
\begin{figure}
\centering
\includegraphics[scale=0.285]{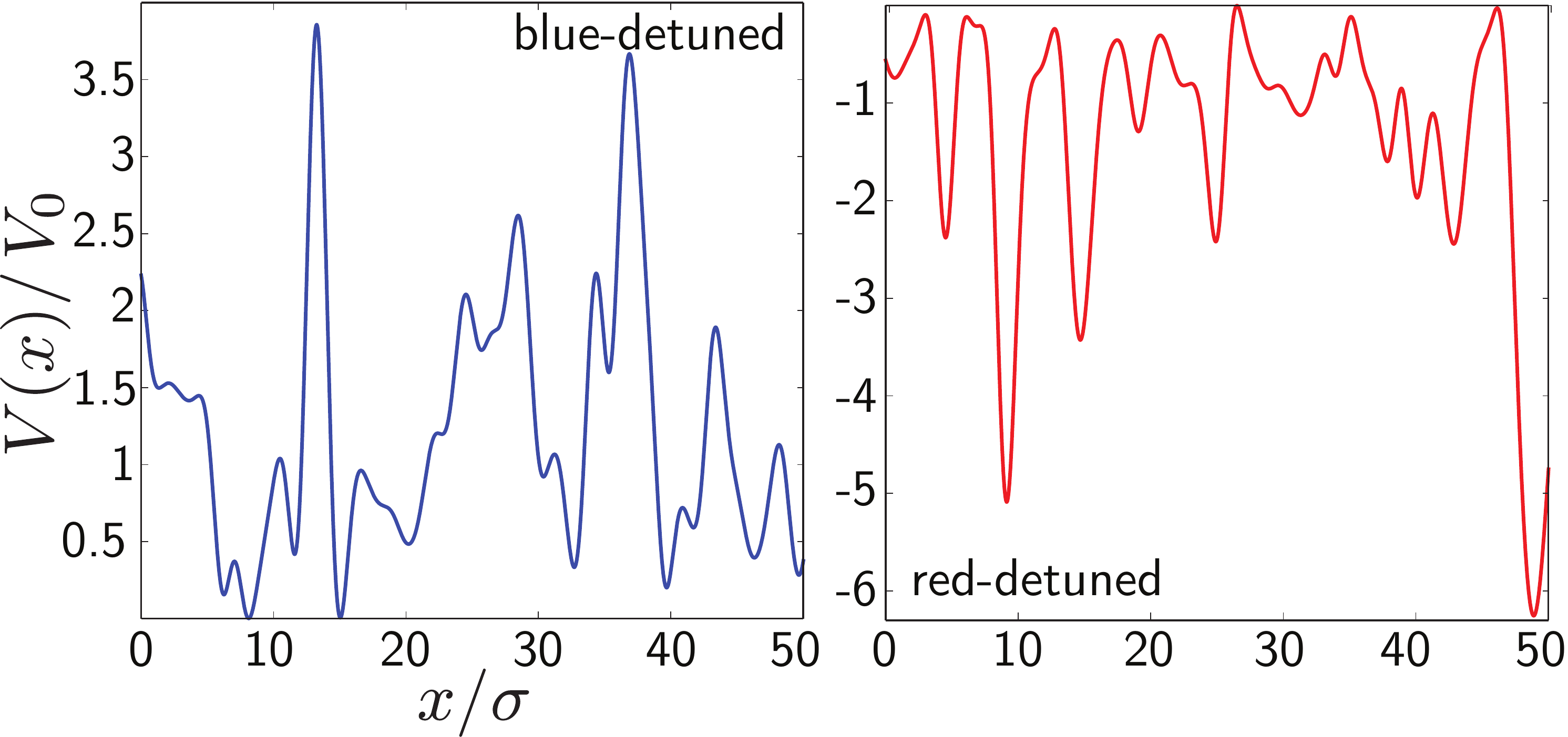}
\caption{(Color online) Numerical realizations of a red- (left) and a blue-detuned (right) 1D speckle potential. The on-site distribution is given by Eq. \eqref{eq-distribspeckleonsite} and the two-point correlation function by Eq. \eqref{eq-specklecorrel}. The procedure used to numerically generate the speckle is explained in the main text.}
\label{fig-speckles}
\end{figure}

We consider a cloud of non-interacting atoms of mass $m$, subjected to a random potential $V(\br)$. Its dynamics is governed by the Hamiltonian
\begin{equation}
H = \frac{\bs{p}^2}{2m} + V(\br),
\end{equation}
where $\bs{p}=-i\hbar\bs{\nabla}$. The coordinate vector $\br \in [0,L]^d$ lies in a $d$-dimensional cubic volume of linear size $L$ that we will eventually make tend to infinity. In the following, averaging over the random potential will be indicated by an overline: $\overline{(\dots)}$. 
In practice, speckle potentials are obtained by transmission or reflection of a laser through a rough plate. The resulting potential $V(\br)$ felt by atoms subjected to this light is proportional to the square of a complex Gaussian field \cite{Goodman08}, with a sign that depends on the laser detuning with respect to the considered two-level transition. This potential has the following on-site distribution:
\begin{equation}
P \left[V(\textbf{r}) \right]=
\frac{1}{V_0}\theta \left[ \pm V(\br) \right] \text{exp} \left[ \mp \frac{V(\br)}{V_0} \right],
\label{eq-distribspeckleonsite}
\end{equation}
where $\theta(\dots)$ is the Heaviside theta function. The disorder strength $V_0>0$ enters both the average $\overline{V(\br)}=\pm V_0$ and the variance $\overline{V(\textbf{r})^2} - \overline{V(\textbf{r})}^2=V_0^2$. In Eq. (\ref{eq-distribspeckleonsite}), the upper sign refers to a blue-detuned speckle potential, bounded by zero from below, and the lower sign to a red-detuned speckle potential, bounded by zero from above. Another quantity that we will frequently encounter in the following is the two-point correlation function $\overline{V(\br)  V(\br')} - \overline{V(\br)}^2$. For the isotropic speckles considered in this paper, the two-point correlation function depends only on $|\br-\br'|$. It decays over a typical distance $\sigma$, referred to as the correlation length \cite{Goodman08}. $\sigma$ defines an important characteristic energy scale, the so-called correlation energy~\cite{Kuhn07}:
\begin{equation}
 E_{\sigma} = \frac{\hbar^2}{m \sigma^2}.
\end{equation}
The two-point correlation function can take various forms depending on the experimental setup \cite{Goodman08}. The approach developed in this paper in principle applies to any shape of the correlation function, but the results for the spectral function and the DoS turn out to very weakly depend of it, provided the proper value of $\sigma$ is chosen. Consequently, for definiteness we will only consider the Gaussian case in the following:
\begin{equation}
\overline{V(\textbf{r})  V(\br')}  - \overline{V(\br)}^2 = V_0^2 \text{exp} \left( - \frac{\vert \br-\br' \vert^2}{2 \sigma^2} \right).
\label{eq-specklecorrel}
\end{equation}
As an example, we show in Fig. \ref{fig-speckles} a numerical disorder realization of both a blue and a red-detuned 1D speckle potential. To generate these realizations, we use a numerical procedure that precisely describes the experimental scenario: we first generate a spatially uncorrelated complex random Gaussian field in Fourier space, simulating the transmission through the rough plate. This field is then multiplied by a proper cut-off function -- that physically describes the shape of the plate -- which we take Gaussian to reproduce the two-point correlation function (\ref{eq-specklecorrel}). Finally, (the opposite of) the modulus square of the field in coordinate space gives the blue-(red-)detuned speckle potential visible in the observation plane \cite{Goodman08}.

\subsection{Definitions, semiclassical regime}

The figure of merit of this paper is the spectral function, defined as
\begin{equation}
\label{Ak_def}
A_\bk(\epsilon) =  \overline{\bra{\bk}  \delta \left( \epsilon - H \right)  \ket{\bk}}. 
\end{equation}
Physically, the spectral function is the probability density for a plane-wave $\ket{\bk}$ to have energy $\epsilon$ in the potential $V(\br)$. At vanishing disorder, the spectral function is a Dirac delta function centered at energy $\hbar^2 k^2 / 2m$. Upon increasing the disorder, this peak acquires a finite width and, at strong disorder, starts to develop intriguing structures that we wish to explore. Introducing the Fourier representation of the Dirac delta function in Eq. (\ref{Ak_def}), it follows that
\begin{equation}
A_{\bk}(\epsilon) = \int_{-\infty}^{\infty} \frac{\text{d} t}{2 \pi \hbar} e^{ i \epsilon t/\hbar} \overline{\bra{\bk} e^{-i H  t/\hbar} \ket{\bk}},
\label{eq-ake-U}
\end{equation}
which establishes the connection with the evolution operator $e^{-iH t/\hbar}$. The spectral function is related to the DoS per unit volume, $\nu(\epsilon),$ through the relation

\begin{equation}
\nu(\epsilon) = \frac{1}{L^d} \text{Tr}\, \overline{ \delta \left( \epsilon - H \right)} = \int \frac{\text{d}^d \bk}{(2\pi)^d}  A_{\bk}(\epsilon).
\label{eq-def-dos}
\end{equation}

There are several energy scales in the problem: $E,E_{\sigma}, V_0$, and only their ratio matter. Of special importance is
the parameter
\begin{equation}
 \eta= \frac{V_0}{E_{\sigma}} = \frac{m\sigma^2V_0}{\hbar^2}.
\label{eq:definition_eta}
 \end{equation}
In this paper, we focus on the so-called \emph{semiclassical regime} characterized by the condition \cite{Trappe15, Falco10}
\begin{equation}
\label{semiclassics_condition}
\eta \gg 1.
\end{equation}
This inequality has a simple interpretation: $\sqrt{\eta}$ is the ratio of the disorder correlation
length $\sigma$ to the de Broglie wavelength of a particle with energy $V_0,$ so that, in the semiclassical regime,
the quantum particle can resolve all the potential fluctuations. Alternatively, a quantum particle with energy$~V_0$ encountering
a potential barrier of height$~V_0$ and thickness$~\sigma$ will have a vanishingly small probability$~\exp(-\sqrt{\eta})$ to tunnel throught it, making the dynamics almost classical.

In the deep semiclassical limit $\eta\to\infty$, the non-commutation between position and momentum can be neglected, so that $\overline{\bra{\bk} e^{-i H  t/\hbar} \ket{\bk}}\approx e^{-i \hbar k^2t/2m}\ \overline{e^{-i V(\br)t/\hbar}}$ and
 Eq. (\ref{eq-ake-U}) yields
\begin{equation}
A_\bk^\text{cl}(\epsilon) = \int_{-\infty}^\infty \frac{\text{d} t}{2\pi\hbar} \frac{e^{i (\epsilon-\epsilon_k) t /\hbar} }{1 \pm i tV_0/\hbar }
= P \left( \epsilon - \epsilon_k \right),
\label{class-ake}
\end{equation}
where $\epsilon_k=\hbar^2\bk^2/(2m)$ and $P \left( \epsilon \right)$ is the on-site potential distribution [Eq. \eqref{eq-distribspeckleonsite}]. In the classical limit, the spectral function thus mimics the on-site distribution (\ref{eq-distribspeckleonsite}) \cite{Trappe15}. With this result in hand, the classical DoS then follows from Eq. (\ref{eq-def-dos}):
\begin{equation}
\nu^{\text{cl}}(\epsilon)= \int_0^{\infty} \text{d}\epsilon_k\nu_0(\epsilon_k)P(\epsilon-\epsilon_k),
\label{eq-def-class-dos}
\end{equation}
where $\nu_0$ is the free-space DoS \cite{Falco10, Trappe15}.

\subsection{Smooth quantum corrections}
\label{subsec-semiclass}

For both the spectral function and the density of states, it is possible to calculate the smooth quantum corrections to the classical limits \eqref{class-ake} and \eqref{eq-def-class-dos} from an analytic expansion in $\hbar$. The calculation of the first quantum correction has been recently carried out in \cite{Trappe15} in the energy domain from Wigner-Weyl formalism \cite{Grammaticos79}. The calculation is also possible in the time domain from an expansion of the evolution operator, as we show in Appendix \hyperref[appendixA]{A}. In any dimension $d$, either of the two approaches leads to
\begin{align}
\begin{split}
&A_\bk(\epsilon) =\int_{-\infty}^\infty \frac{\text{d} t}{2\pi\hbar} \frac{e^{i (\epsilon-\epsilon_k) t /\hbar} }{1 \pm i tV_0/\hbar } \\
& \times \left[ 1 +  \frac{di t^3V_0^2E_\sigma/\hbar^3}{12 ( 1 \pm itV_0/\hbar)}  + \frac{t^4V_0^2E_\sigma/\hbar^4 }{12 (1 \pm itV_0/\hbar)}  \epsilon_k \right] ,
\label{eq-semiclassics-Trappe}
\end{split}
\end{align}
with again the $+$ ($-$) sign for the blue-(red-)detuned speckle. As was noticed in \cite{Trappe15}, Eq. \eqref{eq-semiclassics-Trappe} is correct only at large energies. For $\bk=0$, this can be readily seen from the observation that the first quantum correction term should remain small for the perturbation theory to be valid. This term is of the order of $t^2V_0E_\sigma/\hbar^2\sim (t/\hbar)^2 V_0^2 / \eta$. 
It is useful to define the natural frequency unit in this context:
\begin{equation}
 \omega_0 = \sqrt{\frac{V_0}{m\sigma^2}} = \frac{V_0}{\hbar \sqrt{\eta}},
 \label{eq:omega0}
\end{equation}
which is the typical oscillation frequency in a potential well of height $V_0$ and size $\sigma.$
The condition of validity of Eq.~(\ref{eq-semiclassics-Trappe}) simply reads $\omega_0t\ll 1.$ The Fourier integral over time is then 
well approximated if:
\begin{equation}
\epsilon \gg \hbar \omega_0 = \frac{V_0}{\sqrt{\eta}}.
\end{equation}
If one performs the Fourier integral in Eq. (\ref{eq-semiclassics-Trappe}) at energies smaller than $V_0 / \sqrt{\eta}$, the failure of the perturbation expansion manifests itself as unphysical singularities (delta functions and derivatives). One should then resort to another approach, which is the object of the next section. In fact, as noted in \cite{Trappe15}, for speckle potentials the low-energy region is non trivial. While the classical spectral function, Eq.~(\ref{class-ake}), has a discontinuity at $\epsilon-\epsilon_k,$ the exact spectral function is widely different: for a blue-detuned speckle, it rigorously vanishes below $\epsilon=0$ and, for $k=0$, it rapidly increases between $\epsilon=0$ and $\epsilon\sim V_0/\sqrt{\eta}$. 
These difficulties are absent for Gaussian potentials~\cite{Trappe15}.

\subsection{Treatment of low energies}

\subsubsection{Harmonic-oscillator approximation}
\label{HO_approx}

We now would like to describe the quantum corrections to the classical limit in the low-energy region $\epsilon \sim V_0 / \sqrt{\eta}$ for speckle potentials. For this purpose, we propose an approach inspired of Gutwiller theory \cite{Gutzwiller90}, for which we here sketch the essential ideas. The starting point is the Van Vleck form of the propagator, valid in the semiclassical regime \cite{Tabor83, Haake10}:
\begin{equation}
\overline{\bra{\br} e^{-iHt/\hbar}  \ket{\br'} }
 \simeq \overline{  \sum_{\alpha} 
\left( \dots \right) e^{i S^\alpha(\br,\br',t)/\hbar } },
\label{vanvleck}
\end{equation}
where the sum runs over classical trajectories leading from $\br'$ to $\br$ during the time span $t$. $S^{\alpha}(\br,\br',t)$ is the classical action associated with the classical trajectory $\alpha$. 
We do not give the expression of the prefactors here. Their exact value is not important for the present preliminary discussion, where we remain at a qualitative level and want
only to discuss which classical trajectories give the most important contributions.
The sum over all classical trajectories is a very complicated one, obviously different for each disorder
realization, making the averaging \textit{a priori} rather complex. One can nevertheless envision that the statistical
properties of the potential may have a strong influence. For a blue-detuned speckle at low energy,
there will be essentially short trajectories trapped in the potential wells, so that it is easy to understand that
the peculiar distribution of energy minima will play a crucial role. 

The spectral function is related to the propagator (\ref{vanvleck}) through the relation
\begin{equation}
\label{Ak_from_propag}
A_\bk(\epsilon)=\int\frac{\text{d}t}{2\pi\hbar}\int \frac{\text{d}^d\Delta\br}{L^d} e^{i\epsilon t/\hbar-i\bk\cdot\Delta\br}
\overline{\bra{\br}e^{-iHt/\hbar}\ket{\br'}},
\end{equation}
where $\Delta\br=\br-\br'$. The integral over time can be performed by a stationary phase approximation, which restricts the contributing classical trajectories 
to those with energy $\epsilon$ \cite{Haake10}. At the low energies $\epsilon<\hbar\omega_0$ we are targeting, such classical trajectories lie in potential wells (resp. inverted potential wells) for blue-detuned 
(resp. red-detuned) speckles. 
We propose to approximate these wells by independent harmonic oscillators \cite{footnote1}. Under this approximation, the stationary phase approximation becomes exact so one can simply replace the propagator \eqref{vanvleck} by the known propagator of an harmonic oscillator (resp. inverted harmonic oscillator) \cite{Altland10}.

\subsubsection{Blue-detuned speckle}

Within the harmonic oscillator approximation described above, Eq. (\ref{Ak_from_propag}) simply reduces to a sum of spectral functions of infinitely many random harmonic oscillators $i$ whose minima $V_i$ are centered at $\br_i$. For the case of a 1D, blue-detuned speckle potential, this reads
\begin{equation}
A_{k}(\epsilon) \simeq \frac{1}{L} 
\overline{ \sum_{x_i} 
 \sum_{n=0}^{\infty}  
 \vert \psi_{n}^{i}(k) \vert^2 \delta \left( \epsilon - \epsilon_n^i \right)},
\label{eq-gen-akeblue0}
\end{equation}
where $\psi_n^i(k)$ is the eigenfunction of the 1D $i^\text{th}$ oscillator in $k$ space:
\begin{equation}
\psi_{n}^i(k) = \frac{(2\pi)^{1/2}}{\sqrt{2^{n} n!}} \left( \frac{\hbar}{\pi m \omega_i} \right)^{1/4} e^{-\frac{\hbar k^2 }{2 m \omega_i}+ik x_i } H_{n} \left(\sqrt{\frac{\hbar k^2}{m \omega_i}}  \right),
\nonumber
\end{equation}
normalized according to $\int dk/(2\pi)|\psi^i_n(k)|^2=1$ and with associated eigenenergy $\epsilon_n^i=V_i+\hbar\omega_i(n+1/2)$.

We now make use of the assumption that the harmonic wells are statistically independent, which allows us to take the sum over $x_i$ out of the disorder average. The latter is then over the random frequency $\omega_i$ and the potential minimum $V_i$ of a single oscillator only. By introducing the joint distribution $P(V_i, \omega_i)$ of these two random variables, we rewrite Eq. (\ref{eq-gen-akeblue0}) as
\begin{equation}
A_k(\epsilon) = 
\rho 
\sum_{n=0}^{\infty}
\int dV_i d\omega_i
P(V_i,\omega_i) 
\vert \psi_{n}^i(k) \vert^2 \delta ( \epsilon - \epsilon_n^i ),
\label{eq-gen-akeblue}
\end{equation}
where $\rho$ is the average density of potential minima. Calculation of the distribution $P(V_i,\omega_i)$ will be the object of Sec. \ref{sec1D}.
Note Eq. (\ref{eq-gen-akeblue}) is only justified if a typical harmonic well accomodates many states. In the semiclassical regime (\ref{semiclassics_condition}), this is indeed the case: as the typical frequency of the oscillator will be $\omega_0$ and the typical depth of a potential well $V_0,$ the number of states contained in the well is $\sim V_0/\hbar\omega_0=\sqrt{\eta}\gg1$. 

\subsubsection{Red-detuned speckle}
\label{RDS_method}

For the red-detuned speckle potential, we proceed similarly.
The potentiel wells which can accomodate an harmonic series of bound states have energy minima
typically of the order of $-V_0,$ that is in a range where the classical approximation works well (see below). In contrast,
we are interested in the energy range around $E=0,$ near the maximum allowed potential, and it is the potential
\emph{maxima} which are relevant. We thus make use of an inverted harmonic-oscillator approximation. In this case however, the representation (\ref{eq-gen-akeblue0}) of the spectral function is not convenient due to the continuous nature of the spectrum of the inverted harmonic oscillator \cite{footnote1b}. We therefore prefer to work in the time domain, using formulation (\ref{eq-ake-U}) for the spectral function (written for a 1D speckle): 
\begin{equation}
A_{k}(\epsilon) \simeq \rho
\int_{-\infty}^\infty \frac{\text{d} t}{2 \pi \hbar} 
e^{i\epsilon t/\hbar}
\overline{\bra{k}e^{-iH_\text{IHO} t/\hbar} \ket{k}},
\label{eq-gen-akered0}
\end{equation}
where $H_\text{IHO}=p^2/(2m)-V_i-m \omega_i^2 (x-x_i)^2/2$.
The 1D inverted harmonic-oscillator propagator in $k$ space is given by \cite{Altland10}
\begin{align}
\begin{split}
&\bra{k}e^{-iH_\text{IHO} t/\hbar} \ket{k}= 2 \pi e^{i V_i t/\hbar} \sqrt{\frac{i \hbar}{2 \pi m \omega_i \text{sh}(\omega_i t) }} \\
& \hspace{0.3cm} \times \text{exp} \left\lbrace - \frac{i \hbar k^2}{m \omega_i} \left[ \text{coth}(\omega_i t) - \frac{1}{\text{sh}(\omega_i t)} \right] \right\rbrace .
\end{split}
\label{eq-1DIHO-prop}
\end{align}
The disorder average is then carried out as in Eq. (\ref{eq-gen-akeblue}), by averaging over $V_i$ and $\omega_i$ with the help of the joint distribution $P(V_i,\omega_i)$. 
By ``returning'' the potential $V(x)\to -V(x),$ we are back to the blue-detuned
potential so the joint distribution of the maxima $P_{\mathrm{red}}(V_i,\omega_i)$ is nothing but the joint
distribution $P_{\mathrm{blue}}(V_i,\omega_i)$ for the minima of a blue-detuned speckle. This symmetry also implies that the density of maxima for a red-detuned speckle is equal to the density of
minima $\rho$ for a blue-detuned speckle.

Note that, in principle, one could use the propagator of the harmonic oscillator in the time domain for calculating the spectral function of the blue-detuned speckle potential as well. This approach  turns however inadequate due to the presence of an infinite number of singularities -- when $\omega_i t$ is an integer multiple of $\pi$ --
 arising in the time integral over the propagator.

\section{Statistics of 1D speckle potentials}
\label{sec1D}

\subsection{Joint distribution $P(V_i, \omega_i)$}
\label{Joint_dist_sec}

In this section, we calculate the joint probability distribution $P(V_i,\omega_i)$ discussed above. From here on we drop the subscript $i$ and merely write $P(V,\omega)$ to lighten the notations. We derive it for the blue-detuned speckle potential, for which it corresponds  to the joint probability of minima and potential curvature around minima. 

The distribution $P(V,\omega)$ is closely related to the joint, conditional probability distribution $P(V(x),  V''(x) \vert V'(x) =0,V''(x)>0)$ of $V(x)$ and its second derivative $V''(x)$ given that $V'(x) =0$ and $V''(x)>0$,  that we propose to calculate first. From here on we use the following abbreviated notation for the potential and its derivatives at point $x$:
\begin{align}
&V\equiv V(x),\ \ V_x \equiv V'(x),\ \ V_{xx} \equiv V''(x).
\end{align}
The above distribution follows from
\begin{align}
\begin{split}
\label{PVxxVx_cond}
& P(V, V_{xx} \vert V_x=0,V_{xx} >0) 
=N\times\lim\limits_{ V_x\to 0 } \frac{P(V,V_x,V_{xx})}{P(V_x)}.
\end{split}
\end{align}
The numerical constant $N$ that appears in Eq. \eqref{PVxxVx_cond} stems from the fact that only positive curvatures are selected on the left-hand side,
whereas on the right-hand side all possible values are understood. It will be later determined from the normalization condition. In order to compute the joint distribution $P(V,V_x,V_{xx})$, we follow Goodman \cite{Goodman08} and write the potential as
\begin{equation}
V= \Re(x)^2 + \Im(x)^2.
\end{equation}
Up to a constant multiplicative factor, $\Re(x)$ and $\Im(x)$ respectively describe the real and imaginary part of the laser electric field at point $x$, from which the speckle potential $V$ is built on. As for the potential, we introduce the following short-hand notations
\begin{align}
& \Re\equiv \Re(x),\ \ \Re_x \equiv \Re'(x),\ \ \Re_{xx} \equiv \Re''(x)\nonumber\\
& \Im\equiv \Im(x),\ \ \Im_x \equiv \Im'(x),\ \ \Im_{xx} \equiv \Im''(x).
\end{align}
The motivation for introducing the fields $\Re$ and $\Im$ is that they are independent Gaussian variables with zero mean and equal variance \cite{Goodman08}. Their derivatives are likewise Gaussian, since any linear transformation of a Gaussian retains Gaussian statistics. They also have a zero mean. As a consequence, the six random variables of interest obey the multi-dimensional Gaussian distribution
\begin{align}
\begin{split}
P(\Re,\Im,\Re_x,\Im_x,\Re_{xx},\Im_{xx}) = & \frac{e^{ -\textbf{u}^t C^{-1} \textbf{u}/2}}{8 \pi^3 \sqrt{\text{det}(C)}},
\end{split}
\label{1Ddistrifield}
\end{align}
where $\textbf{u}^t$ is a row vector with entries $(\Re,\Im,\Re_x,\Im_x,\Re_{xx},\Im_{xx})$, and $C$ is the covariance matrix
\begin{equation}
C =
\begin{pmatrix}
   \overline{\Re \Re} & \overline{\Re \Im} & \overline{\Re \Re_x} & \overline{\Re \Im_x} & \overline{\Re \Re_{xx}} & \overline{\Re \Im_{xx}} \\
   \overline{\Im \Re} & \overline{\Im \Im} & \overline{\Im \Re_x} & \overline{\Im \Im_x} & \overline{\Im \Re_{xx}} & \overline{\Im \Im_{xx}} \\
    \overline{\Re_x \Re} & \overline{\Re_x \Im} & \overline{\Re_x \Re_x} & \overline{\Re_x \Im_x} & \overline{\Re_x \Re_{xx}} & \overline{\Re_x \Im_{xx}} 
\\
   \overline{\Im_x \Re} & \overline{\Im_x \Im} & \overline{\Im_x \Re_x} & \overline{\Im_x \Im_x} & \overline{\Im_x \Re_{xx}} & \overline{\Im_x \Im_{xx}}
\\
    \overline{\Re_{xx} \Re} & \overline{\Re_{xx} \Im} & \overline{\Re_{xx} \Re_x} & \overline{\Re_{xx} \Im_x} & \overline{\Re_{xx} \Re_{xx}} & \overline{\Re_{xx} \Im_{xx}} 
\\
   \overline{\Im_{xx} \Re} & \overline{\Im_{xx} \Im} & \overline{\Im_{xx} \Re_x} & \overline{\Im_{xx} \Im_x} & \overline{\Im_{xx} \Re_{xx}} & \overline{\Im_{xx} \Im_{xx}}
\label{covmatrix1}
\end{pmatrix}.\nonumber
\end{equation}
The entries of this matrix can be explicitly calculated for a blue-detuned speckle potential. This yields
\begin{equation}
C =
\begin{pmatrix}
   F(0) & 0 & 0 & 0 & F''(0) & 0 \\
  0 & F(0) & 0 & 0 & 0 & F''(0) \\
    0 & 0 & - F''(0) & 0 & 0 & 0
\\
   0 & 0 & 0 & - F''(0) & 0 & 0
\\
    F''(0) & 0 & 0 & 0 & F^{(4)}(0) & 0
\\
   0 & F''(0) & 0 & 0 & 0 & F^{(4)}(0)
\end{pmatrix}, \nonumber
\end{equation}
where $F(x)$ is related to the two-point correlation function of the potential $V$ through
\begin{equation}
F(x-x') = \frac{1}{2} \sqrt{\overline{V(x)  V(x')}  - \overline{V(x)}^2}.
\label{Fx_def}
\end{equation}
We then introduce in Eq. (\ref{1Ddistrifield}) the change of variables 
\begin{equation}
\Re = \sqrt{V} \cos\theta, \hspace{0.3cm}
\Im = \sqrt{V} \sin\theta,
\end{equation}
from which we calculate the distribution $P(V,\theta,V_x,\theta_x,V_{xx},\theta_{xx})$, with a corresponding Jacobian equal to $1/8$.
By explicitly evaluating the entries of the $C$ matrix for the Gaussian correlation function \eqref{eq-specklecorrel} and 
calculating the remaining integrals over $\theta$, $\theta_x$ and $\theta_{xx}$ with Mathematica \cite{Mathematica}, we find
\begin{align}
\begin{split}
&P(V,V_x,V_{xx}) =  \frac{ \sigma^4}{4 \sqrt{2\pi} V_0^3 V} e^{-\frac{24V+16V_{xx}\sigma^2+(V_x^2-2V V_{xx})^2 \sigma^4/V^3}{16 V_0}} \\
&  \times \left\{ I_{-\frac{1}{4}}\left[ \frac{(V_x^2-2 V V_{xx})^2 \sigma^4}{16 V^3 V_0} \right]+ I_{\frac{1}{4}}\left[ \frac{(V_x^2-2 V V_{xx})^2 \sigma^4}{16 V^3 V_0} \right] \right\}\\
& \times \sqrt{\frac{(-V_x^2+2V V_{xx})V_0}{V}},
\label{piixixx}
\end{split}
\end{align}
where $ I_{1/4}$ and $I_{-1/4}$ are the modified Bessel functions of the first kind. Note that this expression is valid only when $V_x^2 - 2 V V_{xx} < 0$, a condition fulfilled since only minima of the potential are considered \cite{footnote0}.
The distribution $P(V,V_x,V_{xx})$ is regular with respect to the limit $V_x \rightarrow 0$. In Eq. \eqref{PVxxVx_cond}, we can thus take this limit separately in numerator and denominator, reducing the latter to a numerical constant which can be absorbed in the normalization prefactor $N$.
\\
From the joint distribution \eqref{piixixx}, we are now in position to access the probability $P(V,V_{xx}  \vert V_x=0,V_{xx}>0)$ using Eq. \eqref{PVxxVx_cond}. The result is
\begin{align}
\begin{split}
& P(V,V_{xx}  \vert V_x=0,V_{xx}>0)= 
\frac{N \sqrt{V_{xx}}}{V} e^{-\frac{6V^2+4 V V_{xx}\sigma^2+V_{xx}^2 \sigma^4}{4 V_0 V }} \\
&\times \left[ I_{-\frac{1}{4}}\left( \frac{ V_{xx}^2 \sigma^4}{4 V V_0} \right) + I_{\frac{1}{4}}\left( \frac{V_{xx}^2 \sigma^4}{4 V V_0} \right) \right].
\end{split}
\end{align}
By imposing that the distribution is normalized, we find $N=\sigma^5/(2c V_0^{5/2})$, where $c=[\sqrt{3} \Gamma \left(1/4\right) \Gamma \left(5/4\right)-\Gamma \left(-1/4\right) \Gamma
   \left(7/4\right)]/(3^{3/4} \sqrt{2} \pi)  \simeq 1.00685$, which will be replaced by 1 in the following.

The last stage of the calculation consists in connecting $P(V,V_{xx}  \vert V_x=0,V_{xx}>0)$ to the sought for distribution $P(V,\omega)$. This amounts to changing the variables from $V_x = 0$ to $x$ such that $V_x(x)=0$, and from $V_{xx}$ to $\omega$ such that $m\omega^2 = V_{xx}$. The associated Jacobian is $|dV_{x}/dx\times dV_{xx}/d\omega|=2m^{5/2}\omega^3$. We finally infer 
\begin{align}
\begin{split}
&P(V,\omega)=\frac{1}{V\omega_0} \left( \frac{ \omega}{\omega_0} \right)^4
e^{-\frac{3}{2} \left( \frac{V}{V_0} \right)^2-\left( \frac{\omega }{\omega_0} \right)^2- \frac{V_0}{4 V} \left( \frac{ \omega }{\omega_0 } \right)^4}  \\
&\times 
\left[ I_{-\frac{1}{4}}\left( \frac{V_0}{4V} \left( \frac{\omega}{\omega_0} \right)^4 \right)+ I_{\frac{1}{4}}\left( \frac{V_0}{4V} \left( \frac{\omega}{\omega_0} \right)^4 \right) \right].
  \end{split}
  \label{eq-distrib1D}
\end{align}

\begin{figure}
\adjustbox{trim={0.19\width} {0.08\height} {0.10\width} {0.10\height},clip}{\includegraphics[scale=0.55]{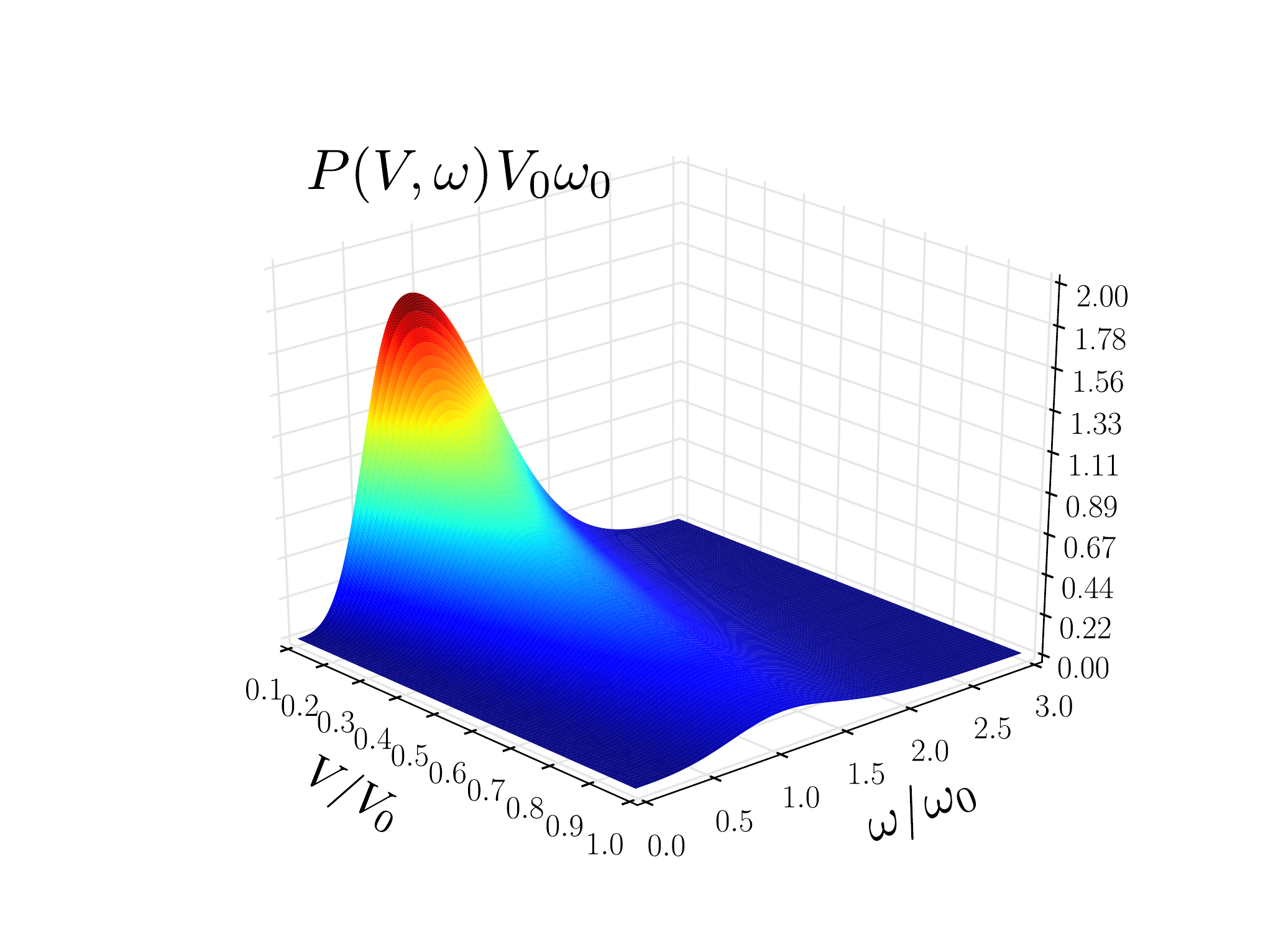}}
\caption{(Color online) Joint distribution $P(V,\omega)$ of minima and potential curvature around minima, for a 1D, blue-detuned speckle potential with Gaussian correlation function [Eq. \eqref{eq-distrib1D}].}
\label{fig-distrib1D}
\end{figure}

The joint distribution is shown in Fig. \ref{fig-distrib1D}. At a given potential minimum $V$, we observe that it is maximum for
$\omega \sim \omega_0$. At smaller $\omega$, the distribution rapidly falls to zero, which supports our description of the speckle potential landscape in terms of purely harmonic wells at low energies.

As we are primilarly interested in low energy minima $V \ll V_0$, it is instructive to express the distribution $P(V,\omega) $ in the limit $V \rightarrow 0$ \cite{Watson66}:
\begin{equation}
    P(V,\omega) V_0 \omega_0 \underset{V \rightarrow 0}{\sim} \sqrt{\frac{2}{\pi } } \sqrt{ \frac{V_0}{V} } \left( \frac{\omega}{\omega_0} \right)^2 e^{-\left( \frac{\omega}{ \omega_0} \right)^2  }.
    \label{eq-distrib1Dasymptotic}
\end{equation}

This asymptotic expression shows that most minima lie at very low $V\ll V_0.$ This phenomenon is ultimately responsible
for the sharp behavior of the spectral function at low energy. A broader distribution of energy minima would smooth out
all peaks and oscillations in the spectral function and DoS, as visible in Fig.~\ref{fig-1Dblueake} and \ref{fig-1Dbluedos}.

\subsection{Density of minima} 

The last unknown quantity is the density of minima $\rho$. To evaluate it, we follow \cite{Weinrib82, Halperin82} and consider the general identity
\begin{equation}
\int \text{d} x \delta ( V'(x) ) f(x) = \sum_n \frac{1}{\vert V''(x_n) \vert} f(x_n),
\end{equation}
valid for any function $f$. The sum is over all point $x_n$ where $V'(x)\equiv V_x$ vanishes. If we choose $f(x)$ to be $\vert V''(x) \vert \equiv\vert V_{xx}\vert$,  then the integral is equal to the number of points at which $V_x$ vanishes. This defines the density of extrema per unit length as
\begin{equation}
\delta ( V_x ) \vert V_{xx} \vert.
\end{equation}
The corresponding density restricted to minima of the potential is
\begin{equation}
\delta ( V_x ) V_{xx} \theta(  V_{xx} ),
\end{equation}
with $\theta$ the Heaviside function. The disorder-averaged density of minima then reads
\begin{equation}
\rho= \int \text{d} V_x \text{d} V_{xx} P(V_x ,V_{xx} ) \delta ( V_x ) V_{xx} \theta( V_{xx} ) .
\end{equation}
Using Eq. \eqref{piixixx}, we obtain $\rho = c'/\sigma$, where $c'\simeq 0.284026$.

\section{1D spectral function and Dos: results}
\label{Sec_results_1D}

\subsection{Spectral function for 1D blue-detuned speckles}

We now evaluate the theoretical prediction (\ref{eq-gen-akeblue}) of the spectral function for 1D, blue-detuned speckle potentials, using Eq. \eqref{eq-distrib1D} for the joint distribution of minima and curvature around minima. By carrying out the integral over $\omega_i$ that ranges from 0 to $\infty$, we find
\begin{equation}
A_k(\epsilon) = 
\frac{c'}{\sigma}
\sum_{n}\int_0^{\epsilon}\text{d} V \frac{\vert \psi_{n}( k ) \vert^2}{\hbar(n+1/2)}     P \left( V,\frac{\epsilon-V}{\hbar (n+1/2 )} \right)
\theta(\epsilon).
\label{eq-1Dblueaek}
\end{equation}
\begin{figure}
\includegraphics[scale=0.45]{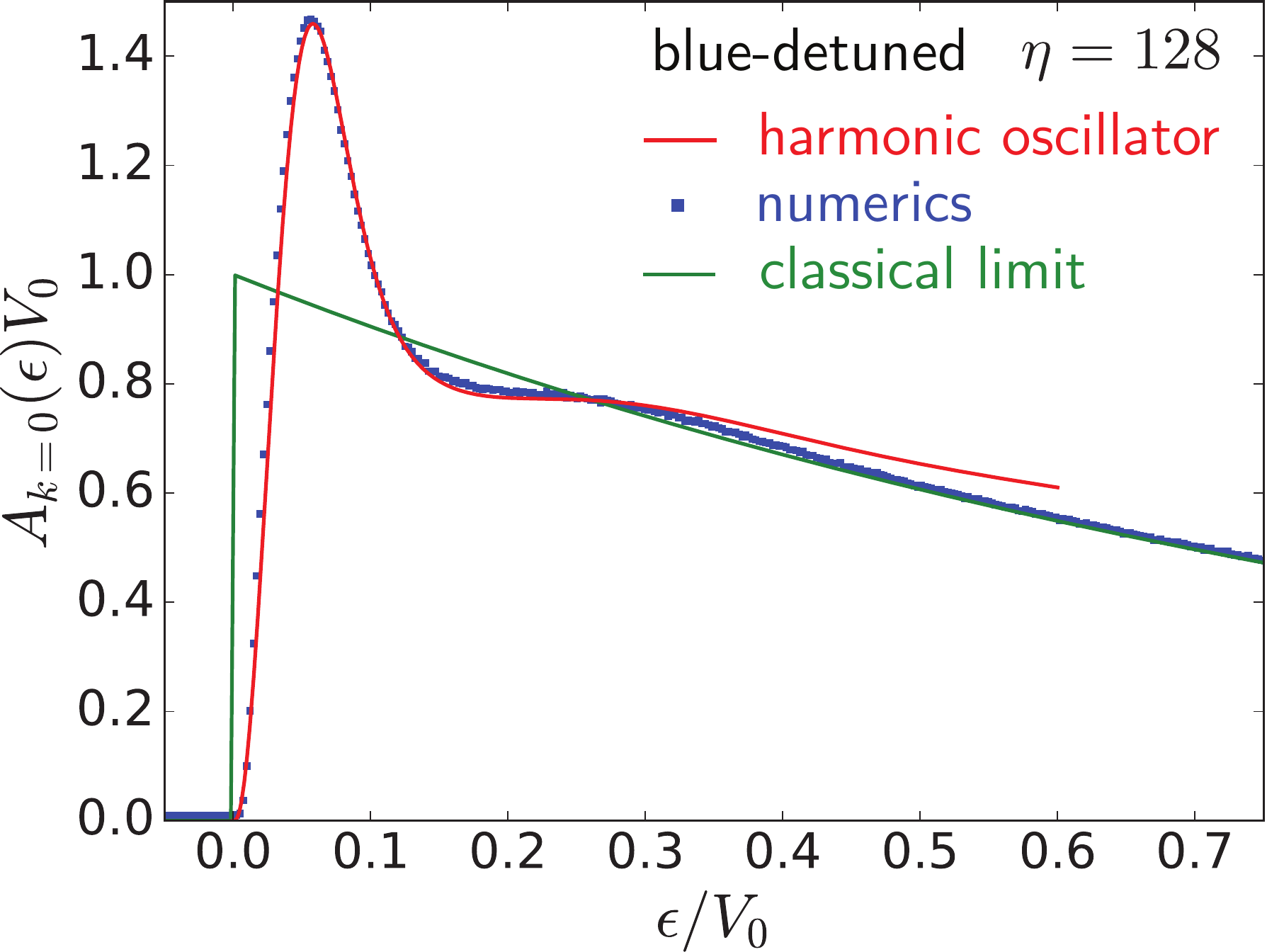}
\caption{(Color online) Spectral function $A_{k=0}(\epsilon)$ as a function of energy in a 1D, blue-detuned speckle potential with Gaussian correlation function, for $\eta=128$. The harmonic-oscillator approximation, Eq. \eqref{eq-1Dblueaek}, is shown as a solid red curve, and the classical limit, Eq. \eqref{class-ake}, as a solid green curve. Blue dots are the result of exact numerical simulations.}
\label{fig-1Dblueake}
\end{figure}
This prediction is shown in Fig. \ref{fig-1Dblueake} as a function of energy, for $k=0$ and $\eta=128$ (solid red curve). As discussed in Sec. \ref{HO_approx}, we expect it to describe low energies. At large energies, the classical limit \eqref{class-ake} (solid green curve in Fig. \ref{fig-1Dblueake}) -- and its smooth quantum corrections (\ref{eq-semiclassics-Trappe}) -- is on the other hand a very good approximation. In order to assess the accuracy of these two limits, we have performed numerical simulations of the spectral function. For these simulations we use a discrete grid of size $L=200 \sigma$ with 4000 grid points and periodic boundary conditions, and compute the spectral function from definition \eqref{eq-ake-U}, using the same approach as described in \cite{Trappe15} to carry out the time evolution. The results are averaged over 50000 disorder realizations, and are shown in Fig. \ref{fig-1Dblueake} as blue dots. We see that the harmonic-oscillator prediction is in excellent agreement with the numerics at low energies. In particular, the high and narrow peak near $\epsilon/V_0\sim 0.05$ and the secondary ``bump'' 
near $\epsilon/V_0\sim 0.25$ are very well described. The peak originates from the ground state of the harmonic oscillator [term $n=0$ in the sum \eqref{eq-1Dblueaek}]; its relatively narrow character originates from the
$\omega$ distribution in Eq.~(\ref{eq-distrib1D}) rather well peaked around $\omega=\omega_0$.
The bump comes from the excited states.

\subsection{Density of states for 1D blue-detuned speckles}
\label{subsec-dosblue}

From definition \eqref{eq-def-dos} and Eq. \eqref{eq-1Dblueaek}, we can compute the DoS for 1D blue-detuned speckles. Carrying out the integral over $k$, we find
\begin{equation}
\nu(\epsilon)  = 
\frac{c'}{\sigma}
\sum_{n}\int_0^{\epsilon}\text{d} V\frac{1}{\hbar(n+1/2)}     P \left( V,\frac{\epsilon-V}{\hbar (n+1/2)} \right)
\theta(\epsilon).
\label{eq-1Dbluedos}
\end{equation}
\begin{figure}
\includegraphics[scale=0.45]{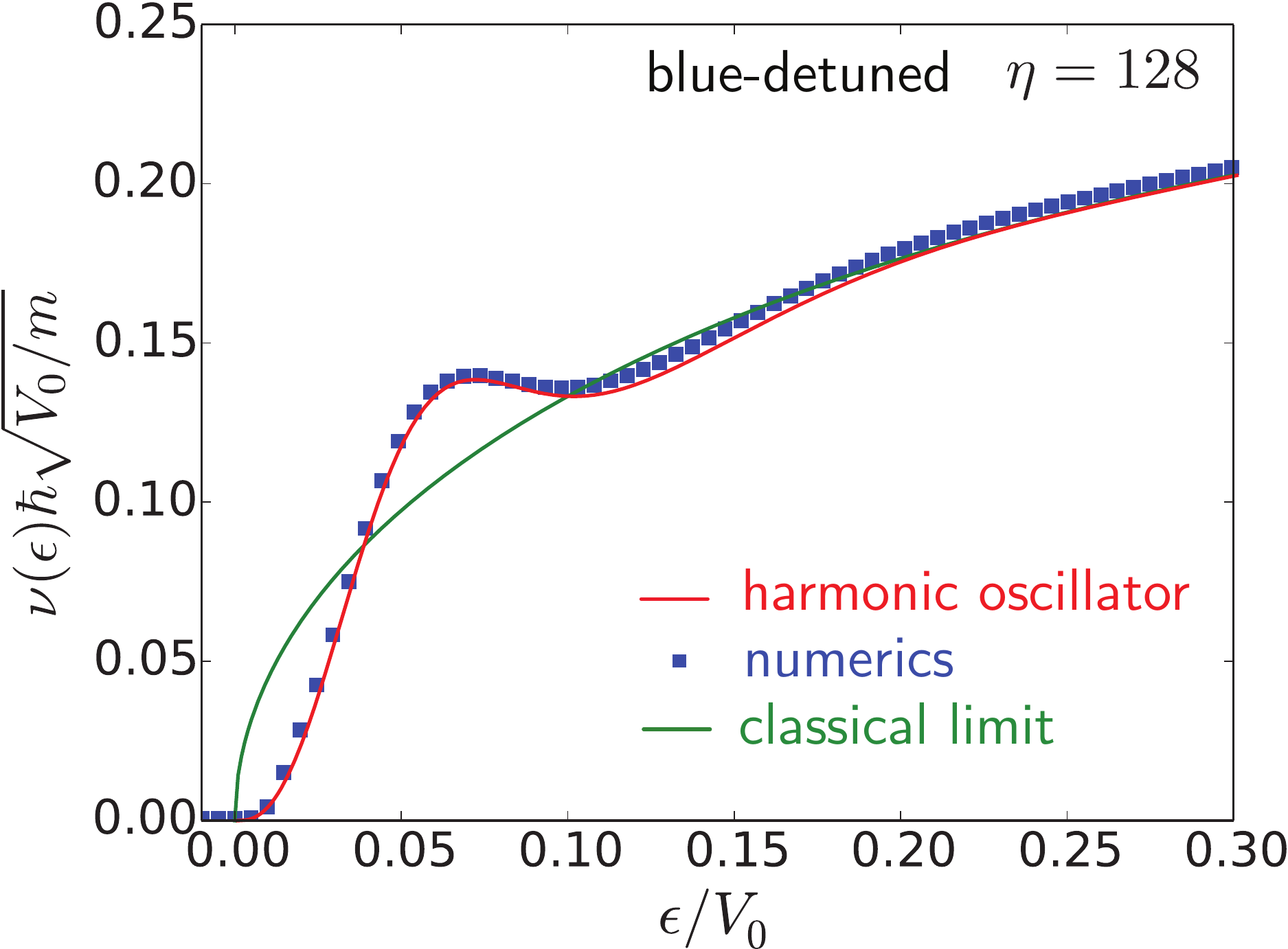}
\caption{(Color online) 
Density of states $\nu(\epsilon)$ as a function of energy in a 1D, blue-detuned speckle potential with Gaussian correlation function. The harmonic oscillator-approximation, Eq. \eqref{eq-1Dbluedos}, is shown as a solid red curve, and the classical limit, Eq. \eqref{eq-def-class-dos}, as a solid green curve. Blue dots are the result of exact numerical simulations.}
\label{fig-1Dbluedos}
\end{figure}
This prediction is shown in Fig. \ref{fig-1Dbluedos} as a function of energy, for $\eta=128$ (solid red curve), together with the classical limit, Eq. (\ref{eq-def-class-dos}) (solid green curve). We have also performed numerical simulations of the DoS, by first computing many spectral functions for $k$ ranging from 0 to $13 \sigma^{-1}$ and then summing over $k$, using a number of grid points between 4000 (at small $k$) and 40000 (for the largest $k$).
These results are shown in Fig. \ref{fig-1Dbluedos} as blue dots. The DoS displays a bump at low energies, which is reminiscent of the narrow peak that shows up in the profiles of the spectral function, see Fig. \ref{fig-1Dblueake}. Indeed, upon increasing $k$ the peak of the spectral function becomes less and less pronounced but remains at the same energy, which results in a smooth bump after summation over $k$. 
As seen in Fig. \ref{fig-1Dbluedos}, at low energies numerical results are very well captured by the harmonic-oscillator prediction. At larger energies $\epsilon>V_0$ (not shown in Fig. \ref{fig-1Dbluedos}), the harmonic-oscillator approximation breaks down and the purely classical limit takes over, eventually leading to $\nu(\epsilon)\simeq \nu_0(\epsilon) =\sqrt{m/(2\epsilon)}/(\pi\hbar)$ for $\epsilon\to\infty$ \cite{Falco10}.

\subsection{Validity of the harmonic-oscillator approximation}
\label{validity_sec}

A simple argument can be used to estimate the energy range where the harmonic-oscillator approximation is valid. 
According to the Virial theorem, equipartition between kinetic and potential energy imposes that $\epsilon_n=m\omega^2 \braket{x^2}_n$ for the mean energy of an eigenstate. 
In order for the speckle potential to be correctly described by a harmonic-oscillator approximation, all states such that $\epsilon_n=\epsilon$ in Eq. \eqref{eq-gen-akeblue0} 
should have an extension $\sqrt{\braket{x^2}_n}$ much smaller than the correlation length $\sigma$, which imposes an upper limit for the energy: 
$\epsilon\ll m\omega^2\sigma^2$ (in case this condition is not fulfilled, anharmonic terms would also come into play). As seen in Sec. \ref{Joint_dist_sec}, 
the most likely value of $\omega$ is $\omega_0$, so the condition becomes
\begin{equation}
\label{validity_OHA}
\epsilon\ll V_0.
\end{equation}
On the other hand, the classical approximation is expected to describe well the spectral function down to energies of order $V_0 / \sqrt{\eta}$ \cite{Trappe15}. 
Therefore, in the region $V_0 / \sqrt{\eta} \ll \epsilon \ll V_0$ both the harmonic-oscillator and the classical approximation provide a good description of the spectral function and of the DoS.

Eq. \eqref{validity_OHA} provides a restriction on the high-energy tail of the spectral function $A_k(\epsilon)$ for the latter to be correctly described by our harmonic-oscillator approximation. A similar argument imposes an additional restriction for the momentum $k$. Indeed, equipartition between kinetic and potential energy for the harmonic oscillator also implies
\begin{equation}
\frac{ \hbar ^2 \braket{k^2}_n}{2m} = \frac{1}{2} m\omega^2 \braket{x^2}_n,
\label{cond_k}
\end{equation}
where $\sqrt{\braket{x^2}_n}$ should be again much smaller than $\sigma$ for the harmonic-oscillator approximation to hold. 
With $\omega\sim \omega_0$, condition \eqref{cond_k} reads
\begin{equation}
\frac{ \hbar ^2 \braket{k^2}_n}{m} \ll V_0 . 
\label{cond_kk}
\end{equation}

The contribution of each eigenstate to the sum in Eq. \eqref{eq-gen-akeblue0} being proportional to $\vert \psi_n(k) \vert^2$, the sum is dominated by eigenstates having $\sqrt{\braket{k^2}_n}$ of the order of $k$, such that criterion \eqref{cond_kk} leads to
\begin{equation}
\epsilon_k \ll V_0.
\end{equation}

In any case, the harmonic oscillator approximation is a good one in the region $\epsilon,\epsilon_k \sim \hbar \omega_0$ where the quantum corrections are important, while the purely classical result (\ref{class-ake}) takes over at higher energy $\epsilon,\epsilon_k \sim
V_0.$

\subsection{Spectral function for 1D red-detuned speckles}

\begin{figure}
\includegraphics[scale=0.45]{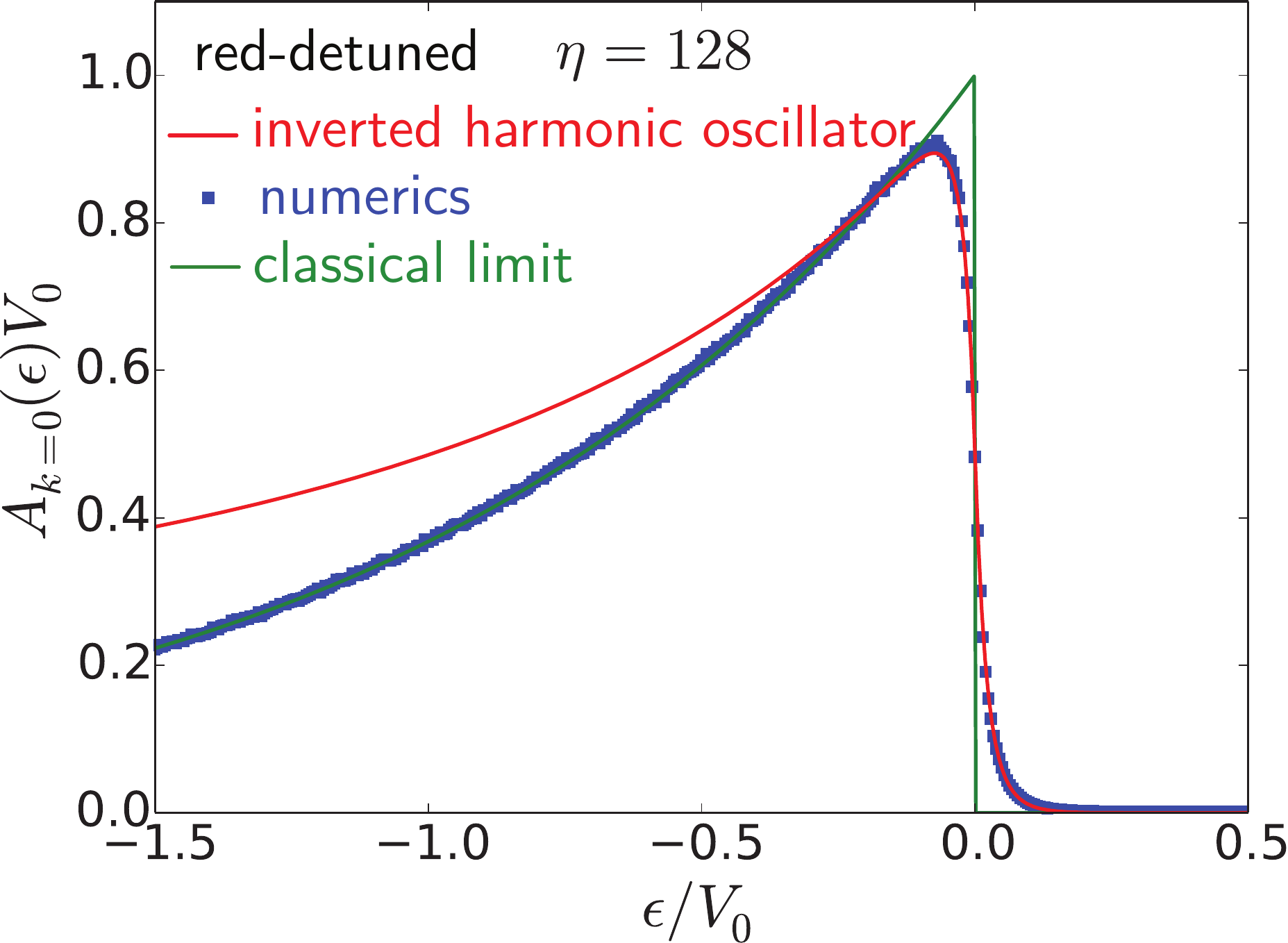}
\caption{(Color online) Spectral function $A_{k=0}(\epsilon)$ as a function of energy in a 1D, red-detuned speckle potential with Gaussian correlation function, for $\eta=128$. The inverted harmonic-oscillator approximation, Eq. \eqref{eq-1Dredaek}, is shown as a solid red curve, and the classical limit, Eq. \eqref{class-ake}, as a solid green curve. Blue dots are the result of exact numerical simulations.}
\label{fig-1Dredake}
\end{figure}

For 1D, red-detuned speckle potentials, we make use of the approach explained in Sec. \ref{RDS_method} to calculate the spectral function. Using Eq. \eqref{eq-gen-akered0} and \eqref{eq-1DIHO-prop} together with the joint distribution \eqref{eq-distrib1D} and carrying out the integral over $V$, we find
\begin{widetext}
\begin{align}
\begin{split}
A_k( \epsilon )= &\frac{(2 \pi)^2 c' \sigma^4 m^{5/2}}{\sqrt{2} V_0^{5/2}}
 \int_{-\infty}^{\infty} \frac{\text{d} t}{2 \pi \hbar} e^{i\epsilon t/\hbar}
\int_0^{\infty} \text{d}\omega
\,\omega^{4} 
  \left[ \ I_{-\frac{1}{4}}^2 \left(\frac{m\omega^2\sigma^2}{2V_0}\sqrt{3 - 2 itV_0/\hbar} \right) - 
   I_{\frac{1}{4}}^2 \left(\frac{m\omega^2\sigma^2}{2V_0}\sqrt{3 - 2 itV_0/\hbar} \right) \right] \\
& \times
\sqrt{\frac{i \hbar}{2 \pi m \omega \text{sh}(\omega t) }}
\text{exp} \left\lbrace - \frac{i  \hbar k^2}{m \omega} \left[ \text{coth}(\omega t) - \frac{1}{\text{sh}(\omega t)} \right]-\frac{m\omega^2\sigma^2}{V_0} \right\rbrace.
\label{eq-1Dredaek}
\end{split}
\end{align}
\end{widetext}
This prediction is shown in Fig. \ref{fig-1Dredake} for $k=0$ (solid red curve), together with the classical limit, Eq. (\ref{class-ake}) (solid green curve). Both limits are compared with the result of numerical simulations (blue dots) that use a discrete grid of size $L=200\sigma$ with 4000 grid points, periodic boundary conditions and 50000 disorder realizations.
As seen in Fig. \ref{fig-1Dredake}, the harmonic-oscillator prediction is in good agreement with the numerical results for energies near 0. At smaller energies ($\epsilon \lesssim - V_0$), the description of the speckle potential in terms of inverse harmonic oscillators becomes poor, while the classical limit provides an excellent approximation.

\subsection{Validity of the inverted harmonic-oscillator approximation}

The breakdown of the inverted harmonic-oscillator approximation at energies $\epsilon \lesssim - V_0$ can be understood from a reasoning on the classical action that appears in Eq. \eqref{vanvleck}. Indeed, for the stationary phase approximation to be valid, the time span $t$ associated with a classical trajectory should be such that the classical action $V_0 t / \hbar$ is large, imposing $t \gg \hbar / V_0$. Energies corresponding to such long times fulfill
\begin{equation}
\abs{ \epsilon } \ll V_0.
\end{equation}
Note that this condition is fully similar to that for blue-detuned speckles, Eq. \eqref{validity_OHA}, though it is here deduced from a slightly different argument. Then, the motion of a classical atom of energy $\epsilon=\epsilon_k-m \omega x^2/2$ describes well the dynamics in a red-detuned speckle as long as the excursion $x^2$ remains much smaller than $\sigma$, namely as long as $\epsilon_k+|\epsilon|\ll m \omega \sigma^2/2$. Since $ \omega\sim \omega_0$ and $|\epsilon|\ll V_0$, this leads to 
\begin{equation}
\epsilon_k \ll V_0,
\end{equation}
which is the same validity condition as for blue-detuned speckles.

\subsection{Density of states for 1D red-detuned speckles}
\label{sec-1D-red-DoS}

\begin{figure}
\includegraphics[scale=0.45]{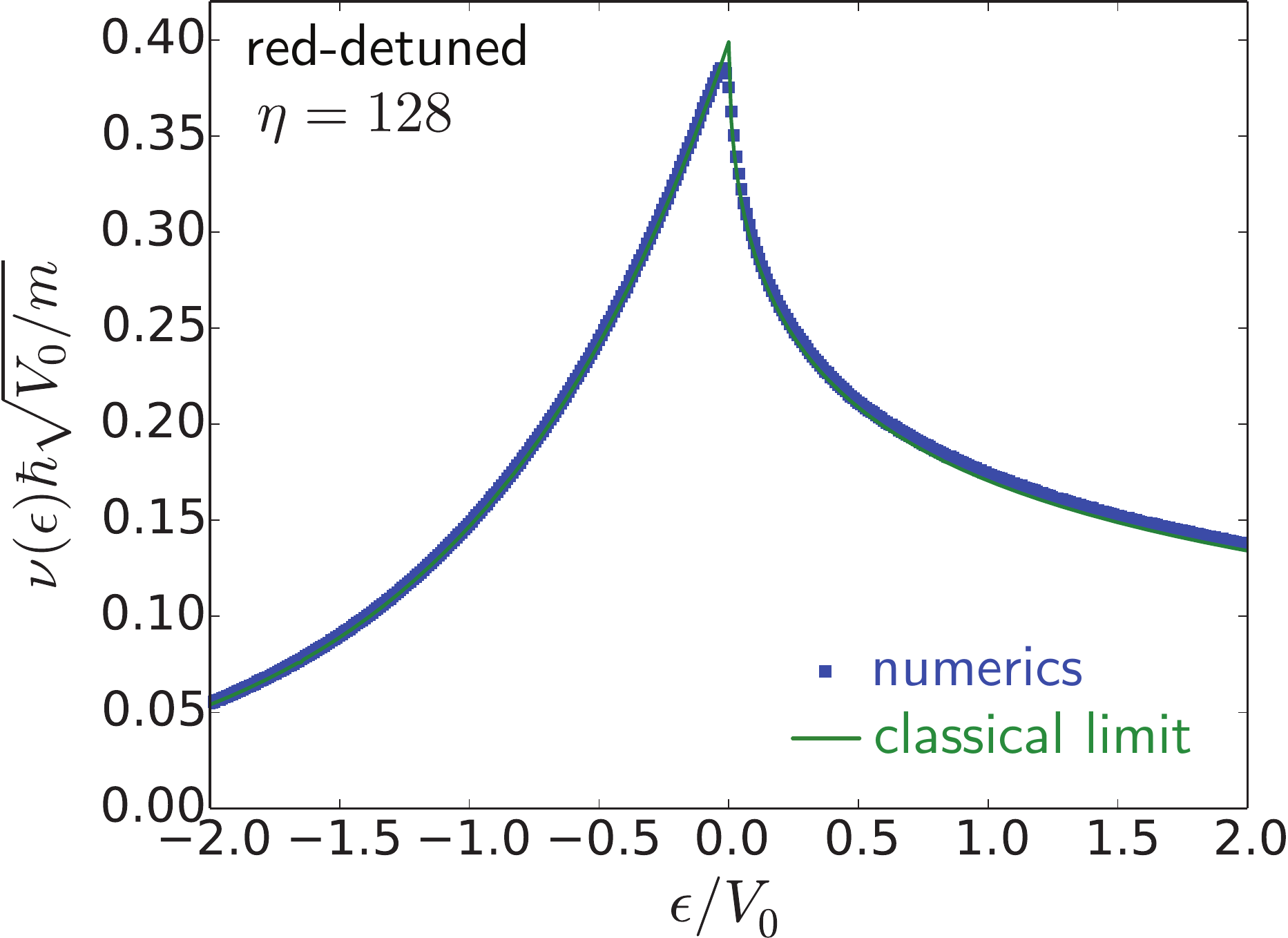}
\caption{(Color online)
Density of states $\nu(\epsilon)$ as a function of energy in a 1D, red-detuned speckle potential with Gaussian correlation function. The classical limit, Eq. \eqref{eq-def-class-dos}, is shown as a solid green curve.  Blue dots are the result of exact numerical simulations.}
\label{fig-1Dreddos}
\end{figure}

We show in Fig. \ref{fig-1Dreddos} as blue dots the DoS in a 1D, red-detuned speckle potential, computed from numerical simulations where we have summed over 208 spectral functions with $k$ ranging from 0 to $13 \sigma^{-1}$, varying the number of grid points from 4000 (for small $k$) to 40000 (for the largest $k$). We also show as the solid green curve the classical prediction \eqref{eq-def-class-dos}. As seen in the figure, the latter already provides an excellent description of the exact results.
This can be understood qualitatively from the Gutzwiller trace formula~\cite{Gutzwiller82,Gutzwiller90} which expresses the density of states as the sum of the classical contribution, Eq.~(\ref{eq-def-class-dos}), and of oscillatory contributions coming from periodic orbits. Around $E=0,$ the periodic orbits in a red-detuned speckle
are long ones with characteristic properties (action, period...) which strongly depend on the disorder realization,
so that all oscillatory contributions cancel out. This is in stark contrast with the blue-detuned speckle
where periodic orbits around $E=0$ are short orbits trapped in the deep potential minima and collectively contribute
to ``bumps'' in the DoS. 

In principle, quantum corrections to the DoS can be obtained from Eq. \eqref{eq-1Dredaek} by evaluating the Fresnel integral over $k$, see Eq. \eqref{eq-def-dos}. The latter can be performed, but the 
remaining integral over $t$ displays an ultraviolet divergence. This divergence already appears in the DoS of the inverted harmonic oscillator, for which it originates of the continuous nature of the spectrum. It thus appears that for the DoS of red-detuned speckles, the description of singular quantum corrections requires to go beyond the inverted harmonic-oscillator approximation, a task that we leave for later work.

\section{Statistics of 2D speckle potentials}
\label{Sec_2D_statistics}

We now turn to the study of 2D speckle potentials which we aim to describe, at low energies, by a 2D harmonic-oscillator approximation. By analogy with the 1D case, we propose to model the speckle potential around an extremum $V(x_i, y_i)$ by a 2D harmonic oscillator (resp. inverted harmonic oscillator) of the form $\pm V \pm m\omega_x^2 (x-x_i)^2/2\pm m\omega_y^2 (y-y_i)^2/2$ with again the $+$ (resp. $-$) sign for blue (resp. red)-detuned speckles, with random frequencies $\omega_x$ and $\omega_y$. Such a description requires the preliminary knowledge of the joint probability distribution $P(V, \omega_x,\omega_y)$ of extrema and potential curvature around extrema. Study of this quantity is the object of the present section. We here focus on  blue-detuned speckle potentials, and then infer the corresponding distribution for red-detuned speckles by the same 
symmetry argument as in one dimension.\\

\subsection{Density of minima at $V=0$}

\begin{figure}
\includegraphics[scale=0.42]{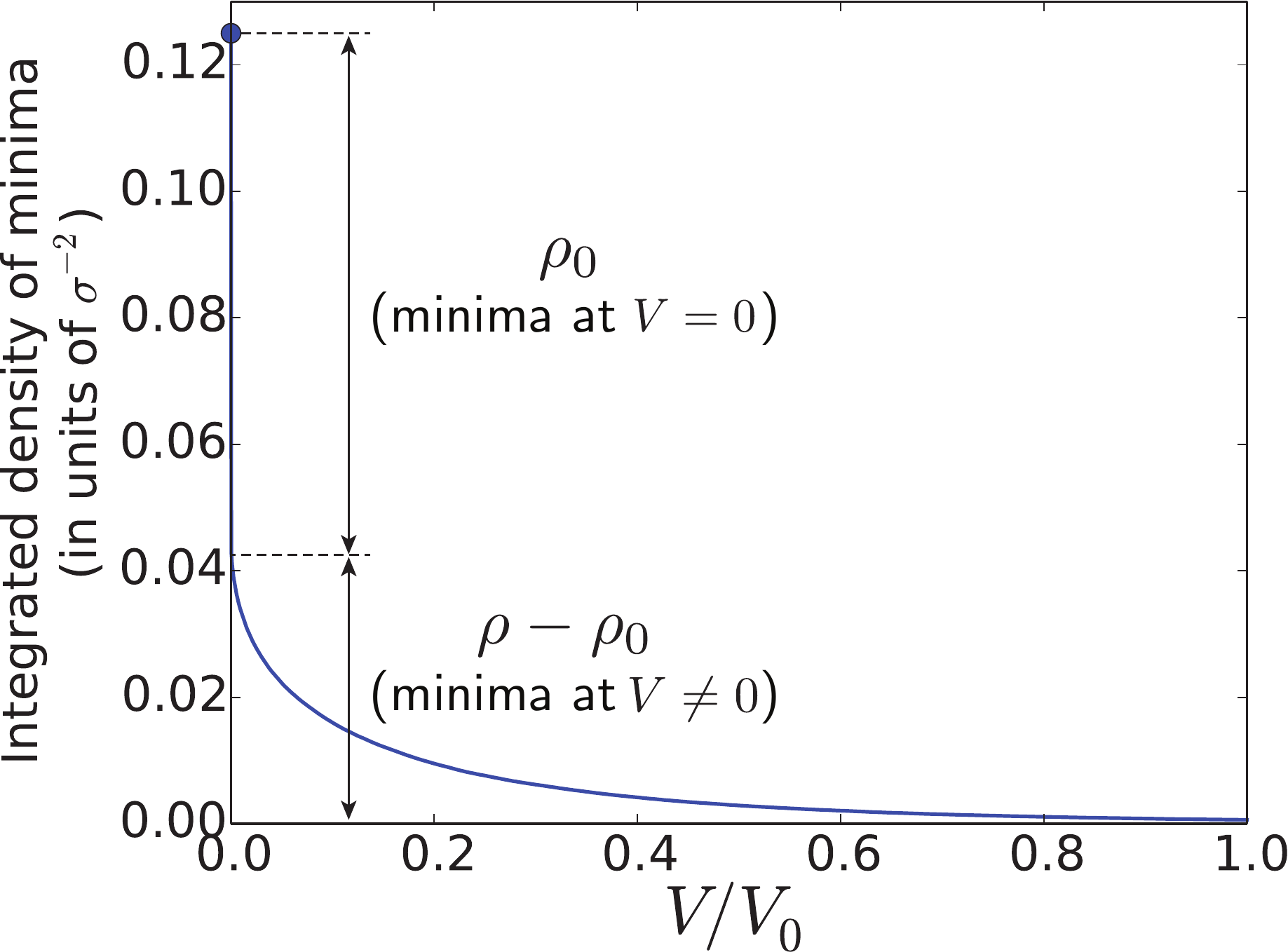}
\caption{(Color online) Blue curve: integrated density of minima (density of minima whose  depth is greater than $V$) for a 2D, blue-detuned speckle potential with Gaussian correlation function. The results have been obtained numerically on a discrete grid of size $L \times L=400\sigma \times 400\sigma$. }
\label{fig-2Ddensmin}
\end{figure}

2D speckle potentials have a important difference with 1D potentials: they present a finite density of points exactly at $V = 0$ \cite{Weinrib82}.
In writing the blue-detuned speckle potential as
\begin{equation}
V(x,y) =\Re(x,y)^2 + \Im(x,y)^2,
\end{equation}
these points are minima that correspond to the intersections of the curves $\Re(x,y) = 0$ and $\Im(x,y)=0$.  
Before considering the distribution $P(V, \omega_x,\omega_y)$, let us first examine the proportion of minima at $V=0$ and at $V \neq 0$. To this end, we have numerically computed the integrated density of minima, i.e. the density of minima whose depth is greater than $V$. To distinguish between minima at $V=0$ and minima at $V\ne 0$, we have exploited the sensitivity (resp. insensitivity) of the minima at $V=0$ (resp. $V\ne 0$) with respect to a change in the spatial discretization (number of grid points).
The results of these simulations are shown in Fig. \ref{fig-2Ddensmin}. They have been obtained on a discrete grid of size $L \times L=400\sigma \times 400\sigma$, by varying the number of grid points between $18000$ and $26000$ along $x$ and $y$. 
The discontinuity of the integrated density of minima at $V=0$ visible in Fig. \ref{fig-2Ddensmin} defines $\rho_0$, the density of minima at $V=0$. We find that approximately $\rho_0/\rho\sim 65 \%$ of all minima lie at $V=0$ \cite{footnote2}. Note that this result is confirmed by an analytical prediction derived in \cite{Weinrib82}:
\begin{equation}
\rho_0 = \left[\frac{-4\pi F(0)}{\boldsymbol{\nabla}^2_\br F(\br)|_{\br=0}} \right]^{-1} ,
\label{2Ddensmin}
\end{equation}
where $F(\br-\br')=\sqrt{\overline{V(\br)  V(\br')}  - \overline{V(\br)}^2}/2$.  For the Gaussian correlation function \eqref{eq-specklecorrel}, this explicitely gives $\rho_0 = 1/(4 \pi\sigma^2)\simeq 0.08/\sigma^2$.

In two dimensions, the majority of minima thus lies at $V=0$. 
To keep the discussion and the calculation as simple as possible, as a first approximation, we keep only the minima at $V=0$
 in the 2D semiclassical description. We will discuss the validity of this approximation in Sec. \ref{sec-2D-blue-aek}. 
 The joint distribution of interest $P(V, \omega_x,\omega_y)$ reduces to
 
\begin{equation}
P(V, \omega_x,\omega_y) \simeq P(\omega_x,\omega_y) \delta(V),
\end{equation}
where $P(\omega_x,\omega_y)$ is the 2D joint distribution of potential curvatures around a minimum $(x_i,y_i)$ where $V(x_i,y_i) = 0$.

\subsection{Joint distribution $P(\omega_x,\omega_y)$}

The distribution $P(\omega_x,\omega_y)$ is closely related to the joint, conditional probability distribution $P(\omega_x,\omega_y|V(x_i,y_i)=0)$ of potential curvatures given that $V(x_i,y_i) = 0$. To calculate this distribution, we first expand $V(x,y)$ up to second order in the vicinity of $(x_i,y_i)$ as
\begin{equation}
V(x,y)\simeq \frac{1}{2} XAX^{t},
\label{eq:quadratic_form}
\end{equation}
where $X = (x-x_i,y-y_i)$ and
\begin{equation}
A = 
\begin{pmatrix}
     \partial_x^2 V(x,y)  &   \partial_x \partial_y V(x,y)  \\
    \partial_y \partial_x V(x,y)   &  \partial_y^2 V(x,y)  \\
\end{pmatrix}.
\end{equation}
By diagonalizing the quadratic form (\ref{eq:quadratic_form}) (which is possible since the matrix $A$ is symmetric), we can describe a well of the speckle potential in terms of two  independent 1D harmonic oscillators, whose curvatures $\omega_x$ and $\omega_y$ are related to the eigenvalues $\lambda_1$ and $\lambda_2$ of $A$ through $\omega_x=\sqrt{\lambda_1/m}$ and $\omega_y=\sqrt{\lambda_2/m}$. The calculation of the joint distribution of the eigenvalues ($\lambda_1,\lambda_2$) is done in Appendix \hyperref[appendixB]{B} for clarity. The corresponding result for $P(\omega_x, \omega_y|V=0)$ is 
\begin{equation}
\label{P2D_cond}
P\left( \omega_x, \omega_y| V=0 \right) = \frac{2}{\omega_0^4} \vert \omega_y^2 - \omega_x^2 \vert e^{-\left( \omega_x^2 + \omega_y^2 \right) / \omega_0^2}.
\end{equation} 
The sought for distribution $P \left( \omega_x, \omega_y \right)$ then follows from the change of variables from $V(x_i,y_i)=0$ to $(x_i,y_i)$ such that $V(x_i,y_i) = 0$: $P \left( \omega_x, \omega_y \right)=(d^2V/dxdy)P\left( \omega_x, \omega_y| V=0 \right)$, where $\text{d}^2 V$ is the change in the surface element defined by the 2D curve $V(x,y)$ when $x$ varies from $x_i$ to $x_i + \text{d} x$ and $y$ varies from $y_i$ to $y_i + \text{d} y$.
Since $V(x,y)\simeq m\omega_x^2(x-x_i)^2/2+m\omega_y^2(y-y_i)^2/2$ in the vicinity of a minimum, we expect this change to be proportional to $\omega_x\omega_y dx dy$, such that

\begin{equation}
P\left( \omega_x, \omega_y \right)\propto \omega_x\omega_y P\left( \omega_x, \omega_y| V=0\right).
\end{equation}

\begin{figure}
\includegraphics[scale=0.47]{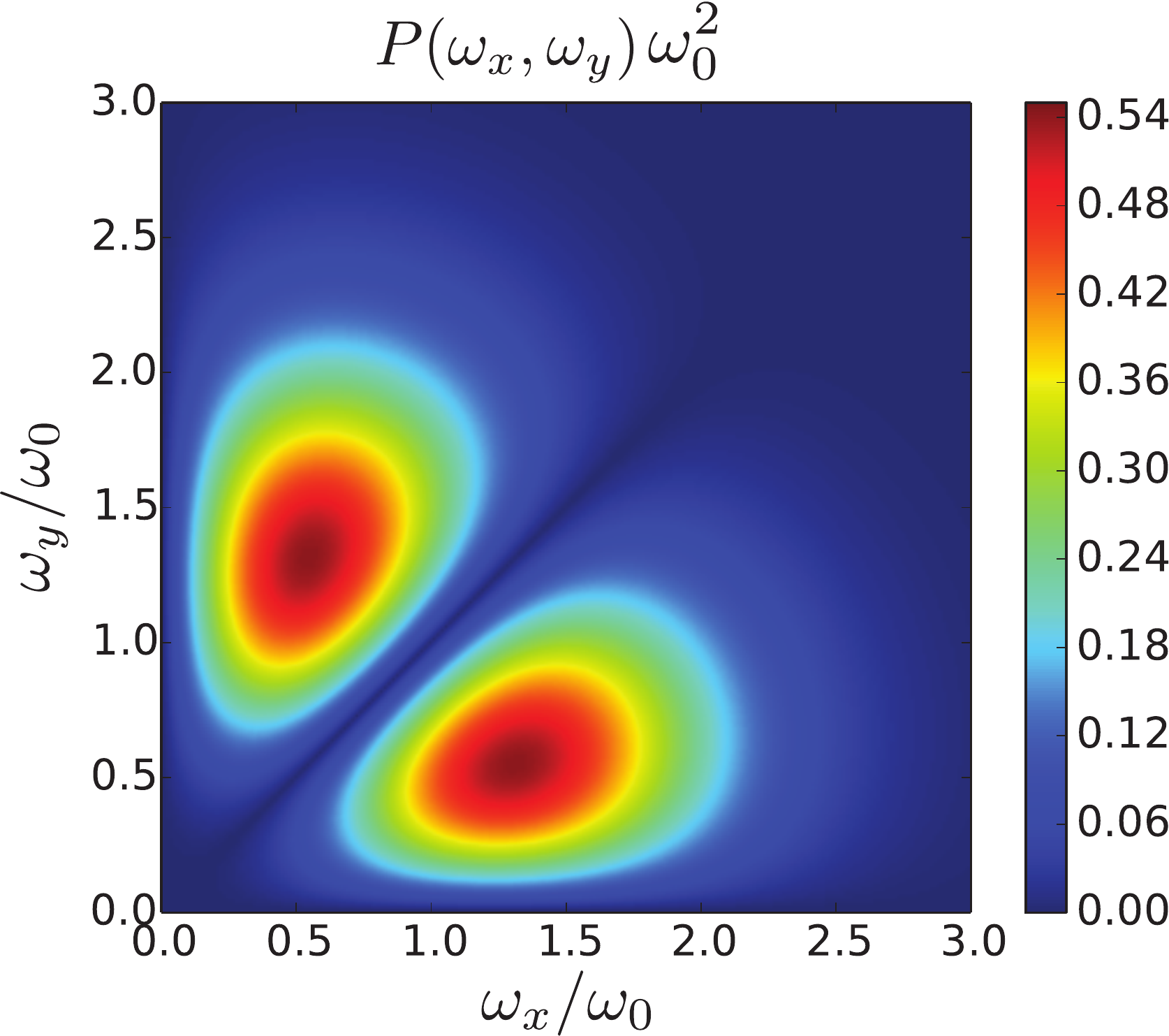}
\caption{(Color online) Density plot of the joint distribution $P(\omega_x,\omega_y)$ at a point  where $V=0$ for a 2D, blue-detuned speckle potential with Gaussian correlation function.
\label{fig-distrib2D}}
\end{figure}

\begin{figure}
\includegraphics[scale=0.47]{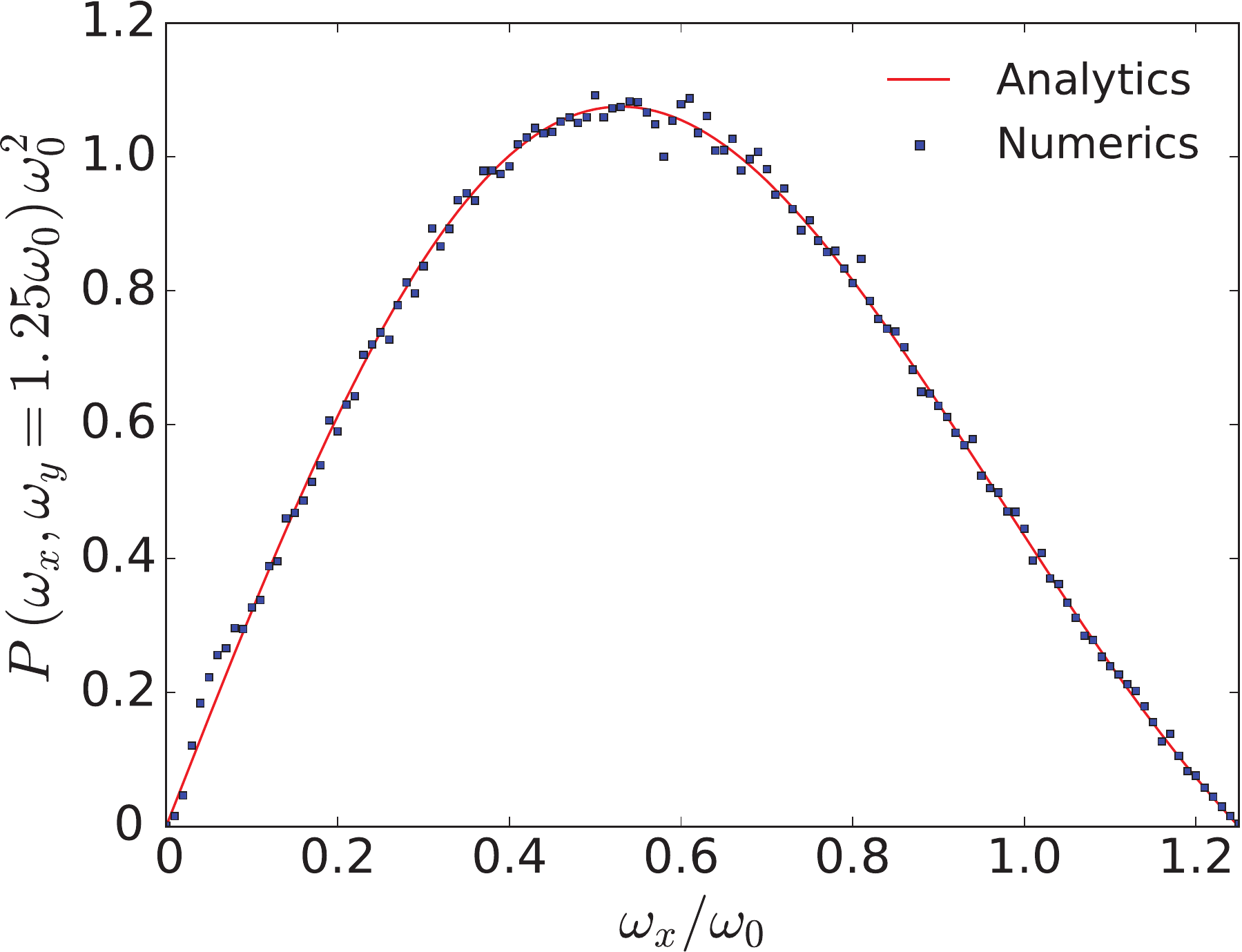}
\caption{(Color online) Cut $P(\omega_x,\omega_y=1.25\omega_0)$ of the joint distribution of curvatures around a minima at $V=0$ for a 2D, blue-detuned speckle potential. Blue dots are the results of numerical simulations and the red curve is Eq. \eqref{eq-distrib2D}.}
\label{cutfig}
\end{figure}

The unknown prefactor is determined from normalization, which eventually leads to
\begin{equation}
P\left( \omega_x, \omega_y \right)=
\frac{4}{\omega_0^6} \omega_x \omega_y  \vert \omega_y^2 - \omega_x^2 \vert e^{-\left( \omega_x^2 + \omega_y^2 \right) / \omega_0^2}.
\label{eq-distrib2D}
\end{equation}
A density plot of $P\left( \omega_x, \omega_y \right)$ is shown in Fig. \ref{fig-distrib2D}. As in one dimension, the distribution rapidly falls to zero at small frequencies, which again supports our description of the speckle potential landscape in terms of purely harmonic wells at low energies.
We have confirmed Eq. \eqref{eq-distrib2D} by numerical simulations of the distribution $P(\omega_x,\omega_y)$, deduced from numerically generated speckle potentials. 
We show in Fig. \ref{cutfig} the numerical cut $P(\omega_x,\omega_y=1.25 \omega_0)$ as a function of $\omega_x$ (blue dots),
together with Eq. \eqref{eq-distrib2D} (red curve), and find a very good agreement.

\section{2D spectral function and Dos: results}
\label{Sec_results_2D}

\subsection{Spectral function for 2D blue-detuned speckles}
\label{sec-2D-blue-aek}

We are now in position to compute the spectral function for 2D, blue-detuned speckle potentials. The 2D counterpart of Eq. (\ref{eq-gen-akeblue}) reads
\begin{eqnarray}
A_\bk(\epsilon)&= &
\rho_0 
\sum_{n_x,n_y=0}^{\infty}
\int \text{d}\omega_x\text{d}\omega_y
P(\omega_x,\omega_y) \nonumber\\
&&\times
\vert \psi_{n_x}(k_x) \vert^2 
\vert \psi_{n_y}(k_y) \vert^2\delta ( \epsilon - \epsilon_{n_x,n_y} ),
\end{eqnarray}
where $P(\omega_x,\omega_y)$ is the joint distribution of curvatures around minima at $V=0$ given by Eq. (\ref{eq-distrib2D}), $\epsilon_{n_x,n_y}=\hbar\omega_x(n_x+1/2)+\hbar\omega_y(n_y+1/2)$ and the eigenfunctions $\psi_{n_x}(k_x)$ are given by Eq. (\ref{eq-gen-akeblue0}) with $n$ replaced by $n_x$ and $k$ replaced by $k_x$, and similarly for $\psi_{n_y}(k_y)$. By performing the integral over $\omega_y$ and using that $\rho_0=1/(4\pi\sigma^2)$, we find
\begin{eqnarray}
A_{\bk}(\epsilon) &= &
\frac{1}{4\pi\sigma^2}
\sum_{n_x,n_y}
\int_0^{\frac{\epsilon}{\hbar(n_x + 1/2)}}\text{d}\omega_x
\frac{\vert \psi_{n_x}( k_x ) \vert^2}{\hbar(n_y + 1/2)}\theta(\epsilon)\nonumber\\
&&\times
\left.\vert \psi_{n_y}(k_y) \vert^2
P\left(\omega_x,\omega_y\right)
\right|_{\omega_y=\frac{\epsilon-\hbar \omega_x(n_x+1/2)}{\hbar(n_y+1/2)}}.
\label{eq-2Dblueaek}
\end{eqnarray}

\begin{figure}
\includegraphics[scale=0.45]{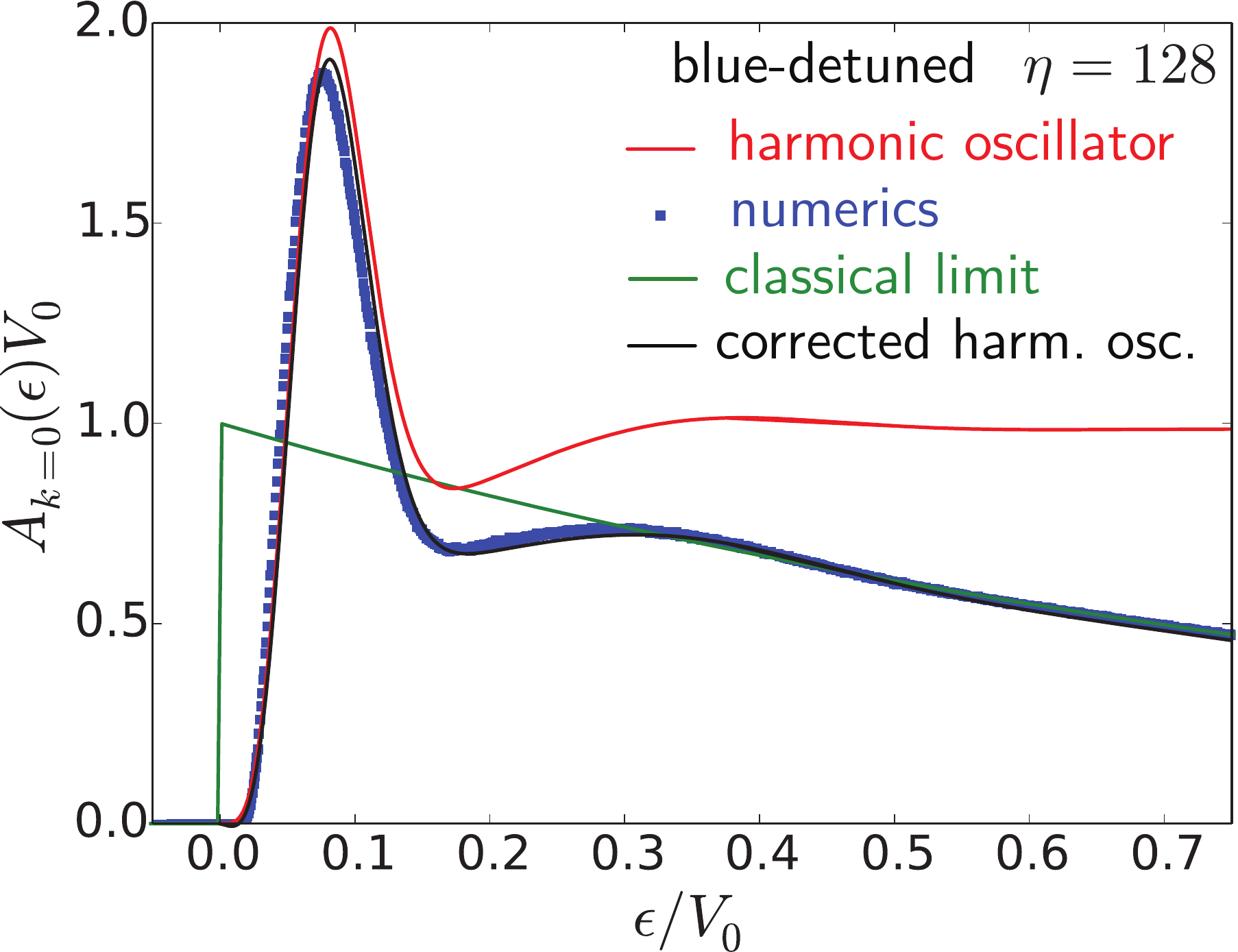}
\caption{(Color online) Spectral function $A_{k=0}(\epsilon)$ as a function of energy in a 2D, blue-detuned speckle potential with Gaussian correlation function, for $\eta=128$. The harmonic-oscillator approximation, Eq. \eqref{eq-2Dblueaek}, is shown as a solid red curve, and the classical limit, Eq. \eqref{class-ake}, as a solid green curve. The corrected harmonic-oscillator approximation, Eq. \eqref{eq-2Dblueaek-cor}, is shown as a solid black curve. Blue dots are the results of exact numerical simulations.}
\label{fig-2Dblueake}
\end{figure}

This prediction is shown in Fig. \ref{fig-2Dblueake} as a function of energy, for $k=0$ and $\eta=128$ (solid red curve). The classical limit \eqref{class-ake}, expected to describe large energies, is also shown as a solid green curve. These results are compared to numerical simulations of the spectral function (blue dots) which use a system size $L \times L= (20 \pi \sigma)^2$ with 600 grid points along $x$ and $y$, periodic boundary conditions and 40000 disorder realizations.
Several observations can de made. Like in one dimension, the harmonic approximation quantitatively
describes the spectral function for energies $\sim V_0/\sqrt{\eta}=\hbar \omega_0.$ The large peak  
is at an energy about twice larger than in one dimension -- compare with Fig.~\ref{fig-1Dblueake} -- because it is the ground state energy of a 2D (instead of 1D) harmonic oscillator. It is also slightly higher and the minimum around $\epsilon/V_0=0.2$
as well as the second bump above are slightly more visible than in one dimension. This is because most potential minima
are exactly at $V=0$ in two dimensions, while this is not true in one dimension, so that an additional smoothing takes place
in the latter case. This must however be taken with a grain of salt: the 2D low-energy peak of the spectral function is not entirely controlled by the ground-state of the harmonic oscillator: excited states also contribute for roughly $25 \%$ of the peak height.
As seen in Fig. \ref{fig-2Dblueake}, deviations of the harmonic-oscillator prediction from the numerical result occur at smaller energy than in one dimension. This phenomenon can be understood from the expression of the spectral function in terms of the propagator of the 2D harmonic oscillator:
\begin{equation}
A_{\bk}(\epsilon) = \rho_0
\int_{-\infty}^\infty \frac{\text{d} t}{2 \pi \hbar} 
e^{i\epsilon t/\hbar}
\overline{\bra{\bk}e^{-iH_\text{HO} t/\hbar} \ket{\bk}},
\label{eq-2Dblue-ake-prop}
\end{equation}
where $H_\text{HO}=\bs{p}^2/(2m)+m \omega_x^2 x^2/2+m \omega_y^2 y^2/2$.
In two dimensions, the propagator $\bra{\bk}e^{-iH_\text{HO} t/\hbar} \ket{\bk}\propto 1/t$ at short times \cite{Altland10}. This singularity is more pronounced than in one dimension where the propagator diverges as $1/\sqrt{t}$. In two dimensions there is thus more weight on short times, which are by 
construction not well captured by the harmonic-oscillator approximation. On the other hand, we know that short times are fairly well described by the classical limit, Eq. \eqref{class-ake}. To improve on the quality of 
the harmonic-oscillator description, we thus propose to replace  the contribution from the pole at $t=0$ by the classical contribution. The contribution from this pole is simple to calculate from Eq. \eqref{eq-2Dblue-ake-prop}: we find $\theta(\epsilon)/V_0$. The classical contribution is given in Eq. \eqref{class-ake}. The above prescription thus leads to
\begin{equation}
A_{\bk}^\text{corr}(\epsilon) \simeq A_\bk(\epsilon) - \frac{\theta(\epsilon)}{V_0} + \frac{\theta(\epsilon)}{V_0} \text{exp} \left( - \frac{\epsilon - \epsilon_k}{V_0} \right),
\label{eq-2Dblueaek-cor}
\end{equation}
where $A_{\bk}(\epsilon)$ is the prediction of the harmonic-oscillator description, Eq. \eqref{eq-2Dblueaek}.
Eq. \eqref{eq-2Dblueaek-cor} is shown in Fig. \ref{fig-2Dblueake} as a solid black curve, and is in very good agreement with the numerical simulations.

The excellent agreement with the numerical calculations justifies \textit{a posteriori} the approximation of keeping only the minima at $V= 0$. 
Such an agreement may surprise the attentive reader as approximately $35 \%$ of the minima have been left aside. The reason for it lies in 
two mechanisms reducing the contribution to the spectral function of minima at $V \ne 0 $ as compared to minima at $V=0$.
First, among the $35 \%$ of minima at $V \ne 0$, only a fraction contributes to the spectral function: as we are interested in very low energies ($\epsilon \ll V_0$), we should keep only the harmonic wells with
associated minimum smaller than $\epsilon$. Second, the smoothing due to the dispersion in $V$ -- compare the 1D oscillations in Fig. \ref{fig-1Dblueake} 
with such a dispersion and the 2D oscillations in Fig. \ref{fig-2Dblueake} where the dispersion is absent -- 
makes the contribution of minima at $V \neq 0$ negligible after application of the corrected harmonic-oscillator prescription [Eq. \eqref{eq-2Dblueaek-cor}].

\subsection{Density of states for 2D blue-detuned speckles}

From definition \eqref{eq-def-dos} and Eq. \eqref{eq-2Dblueaek}, we can compute the DoS for 2D blue-detuned speckles. Carrying out the integral over $k$, we readily find
\begin{eqnarray}
\nu(\epsilon) &=& \frac{1}{4 \pi \sigma^2}
\sum_{n_x,n_y} 
\int_0^{\frac{\epsilon}{\hbar(n_x + 1/2)}}\text{d}\omega_x
\frac{\theta(\epsilon)}{\hbar(n_y+1/2)} \nonumber\\
&&\times   P\left( \omega_x,\frac{\epsilon-\hbar\omega_x \left( n_x + 1/2 \right) }{\hbar(n_y + 1/2)}  \right).
\label{eq-2Dbluedos}
\end{eqnarray}
\begin{figure}
\includegraphics[scale=0.45]{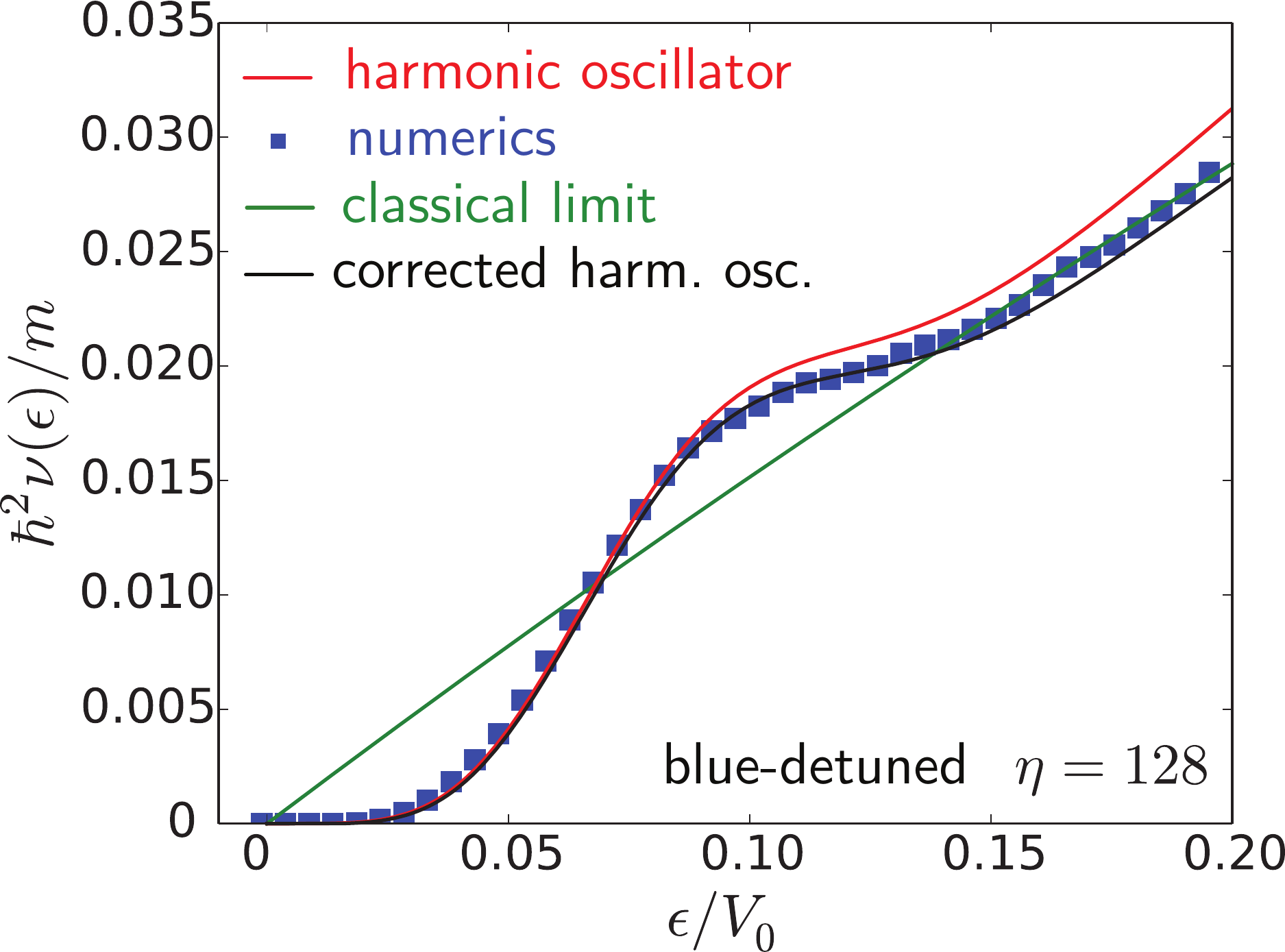}
\caption{(Color online) Density of states $\nu(\epsilon)$ as a function of energy in a 2D, blue-detuned speckle potential with Gaussian correlation function. The harmonic-oscillator approximation, Eq. \eqref{eq-2Dbluedos}, is shown as a solid red curve, and the classical limit, Eq. \eqref{eq-def-class-dos}, as a solid green curve. The corrected harmonic-oscillator description, Eq. \eqref{eq-2Dbluedos-cor}, is shown as a solid black curve. Blue dots are the result of exact numerical simulations.}
\label{fig-2Dbluedos}
\end{figure}
This prediction is shown in Fig. \ref{fig-2Dbluedos} as a function of energy, for $\eta=128$ (solid red curve), together with the classical limit, Eq. (\ref{eq-def-class-dos}) (solid green curve). We have also performed numerical simulations of the DoS. In two dimensions however, the strategy of numerically computing first spectral functions at different $\bk$ and then summing of $\bk$ is numerically demanding. We have thus used a different scheme  that consists in expressing the trace in Eq. \eqref{eq-def-dos} in real space rather than in momentum space:
\begin{equation}
\nu(\epsilon) = \frac{1}{L^2} \text{Tr}\, \overline{ \delta \left( \epsilon - H \right) } = \frac{1}{L^2} \int \text{d}^2 \br \bra{\br} \overline{\delta \left( \epsilon - H \right)  } \ket{\br}.
\end{equation}
The system being translation invariant on average, the integrand is in fact independent of $\br$ so
\begin{eqnarray}
\nu(\epsilon) &= &  \bra{\br=0}  \overline{\delta \left( \epsilon - H \right) }  \ket{\br=0}\nonumber\\
 &=&  \int_{-\infty}^{\infty} \frac{\text{d} t}{2 \pi \hbar} e^{i \epsilon t/\hbar}  \overline{ \bra{0}e^{-i H  t/\hbar}  \ket{0}}.
\label{eq-def-dos-x}
\end{eqnarray}
From Eq. \eqref{eq-def-dos-x}, it thus appears that the DoS can be obtained by numerically propagating a particle initially located at the origin, then recording the value of the wave function at the origin for many different times $t$, and finally taking the Fourier transform with respect to time and averaging over disorder. We have applied this strategy for a system size $L \times L = (10 \pi \sigma)^2$ with $400$ grid points along $x$ and $y$ and 40000 disorder realizations. Results are shown in Fig. \ref{fig-2Dbluedos} as blue dots.
As for the 2D spectral function, we observe deviations of the theoretical prediction \eqref{eq-def-dos-x} from the numerical results at relatively small energies due to a pole $\propto 1/t^2$ in the propagator in \eqref{eq-def-dos-x}. We again correct them by replacing the contribution of this pole by the classical result \eqref{eq-def-class-dos}. This gives
\begin{equation}
\nu^\text{cor}(\epsilon) = 
\nu (\epsilon)-  \frac{m\epsilon \theta(\epsilon)}{2 \pi \hbar^2 V_0} + \frac{m\theta(\epsilon)}{2 \pi \hbar^2}  \left(1 - e^{-\epsilon/V_0} \right).
\label{eq-2Dbluedos-cor}
\end{equation}
This prediction is plotted in Fig. \ref{fig-2Dbluedos} (solid black curve), and describes very well the exact numerical results. 

\subsection{Spectral function for 2D red-detuned speckles}

To evaluate the spectral function for 2D, red-detuned speckle potentials, we proceed as in one dimension and write
\begin{equation}
A_{\bk}(\epsilon) \simeq \rho_0
\int_{-\infty}^\infty \frac{\text{d} t}{2 \pi \hbar} 
e^{i\epsilon t/\hbar}
\overline{\bra{\bk}e^{-iH_\text{IHO} t/\hbar} \ket{\bk}},
\end{equation}
where $H_\text{IHO}=\bs{p}^2/(2m)-m \omega_x^2 x^2/2-m \omega_y^2 y^2/2$.
Making the average over disorder explicit, we have
\begin{eqnarray}
A_{\bk}( \epsilon ) &= & \frac{1}{4\pi\sigma^2} \int_{-\infty}^{\infty} \frac{\text{d} t}{2 \pi \hbar} e^{ i \epsilon t/\hbar} \int_0^{\infty} \text{d} \omega_x \text{d} \omega_y    P\left( \omega_x,\omega_y \right)
\nonumber\\ 
&& \times  \bra{k_x} e^{-i\left[p_x^2/(2m) - m\omega_x^2 x^2/2 \right] t/\hbar} \ket{k_x}
\nonumber\\ 
&& \times \bra{k_y} e^{-i\left[p_y^2/(2m) - m\omega_y^2 y^2/2 \right] t /\hbar} \ket{k_y},
\label{eq-2Dredaek}
\end{eqnarray}
where the 1D inverted harmonic-oscillator propagator is given by Eq. \eqref{eq-1DIHO-prop}. 
\begin{figure}
\includegraphics[scale=0.45]{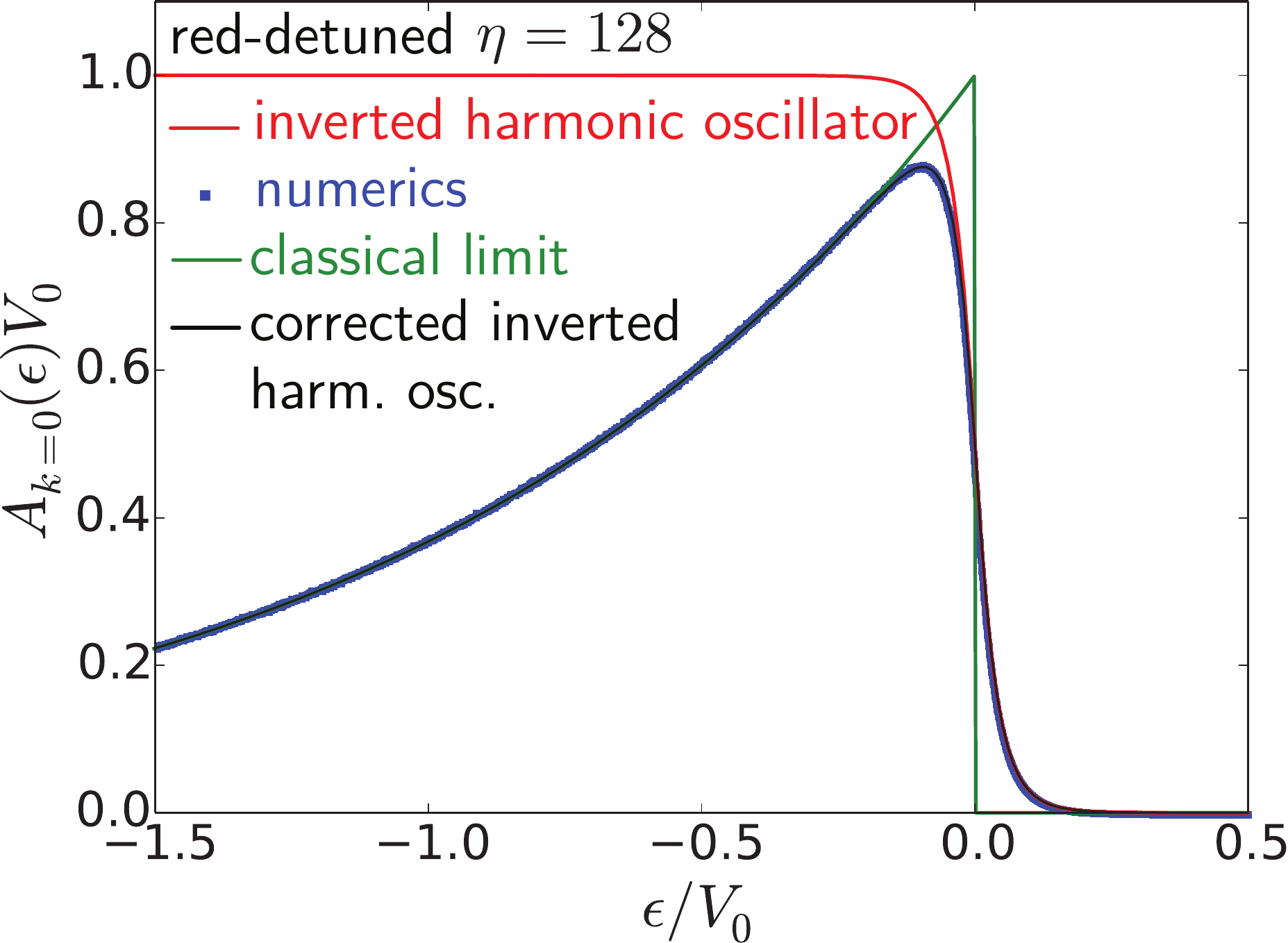}
\caption{(Color online) Spectral function $A_{k=0}(\epsilon)$ as a function of energy in a 2D, red-detuned speckle potential with Gaussian correlation function, for $\eta=128$. The inverted harmonic-oscillator approximation, Eq. \eqref{eq-2Dredaek}, is shown as a solid red curve, and the classical limit, Eq. \eqref{class-ake}, as a solid green curve. 
The corrected inverted harmonic-oscillator approximation, Eq. \eqref{eq-2Dredaek-cor}, is shown as a solid black curve. Blue dots are the results of exact numerical simulations.}
\label{fig-2Dredake}
\end{figure}

Eq. \eqref{eq-2Dredaek} is shown in Fig. \ref{fig-2Dredake} (solid red curve), together with the classical limit, Eq. \eqref{class-ake} (solid green curve). Results of numerical simulations that use a system size $L \times L = (20 \pi \sigma)^2$ with $600$ grid points along $x$ and $y$ and 40000 disorder realizations are also shown (blue dots). As for the blue-detuned speckle, the pole $1/t$ in the propagator gives rise to deviations of the oscillator description from the exact numerical results that are more significant than in one dimension. We again cure them by replacing the contribution of the pole by the classical limit:
\begin{equation}
A_{\bk}^{\text{cor}}(\epsilon) = A_{\bk}(\epsilon) - \frac{\theta(-\epsilon)}{V_0} + \frac{\theta(-\epsilon)}{V_0} e^{(\epsilon -\epsilon_k)/V_0}.
\label{eq-2Dredaek-cor}
\end{equation}
This prediction is plotted in Fig. \ref{fig-2Dredake} (solid black curve), and describes very well the exact numerical results.

\subsection{Density of states for 2D red-detuned speckles}

\begin{figure}
\includegraphics[scale=0.45]{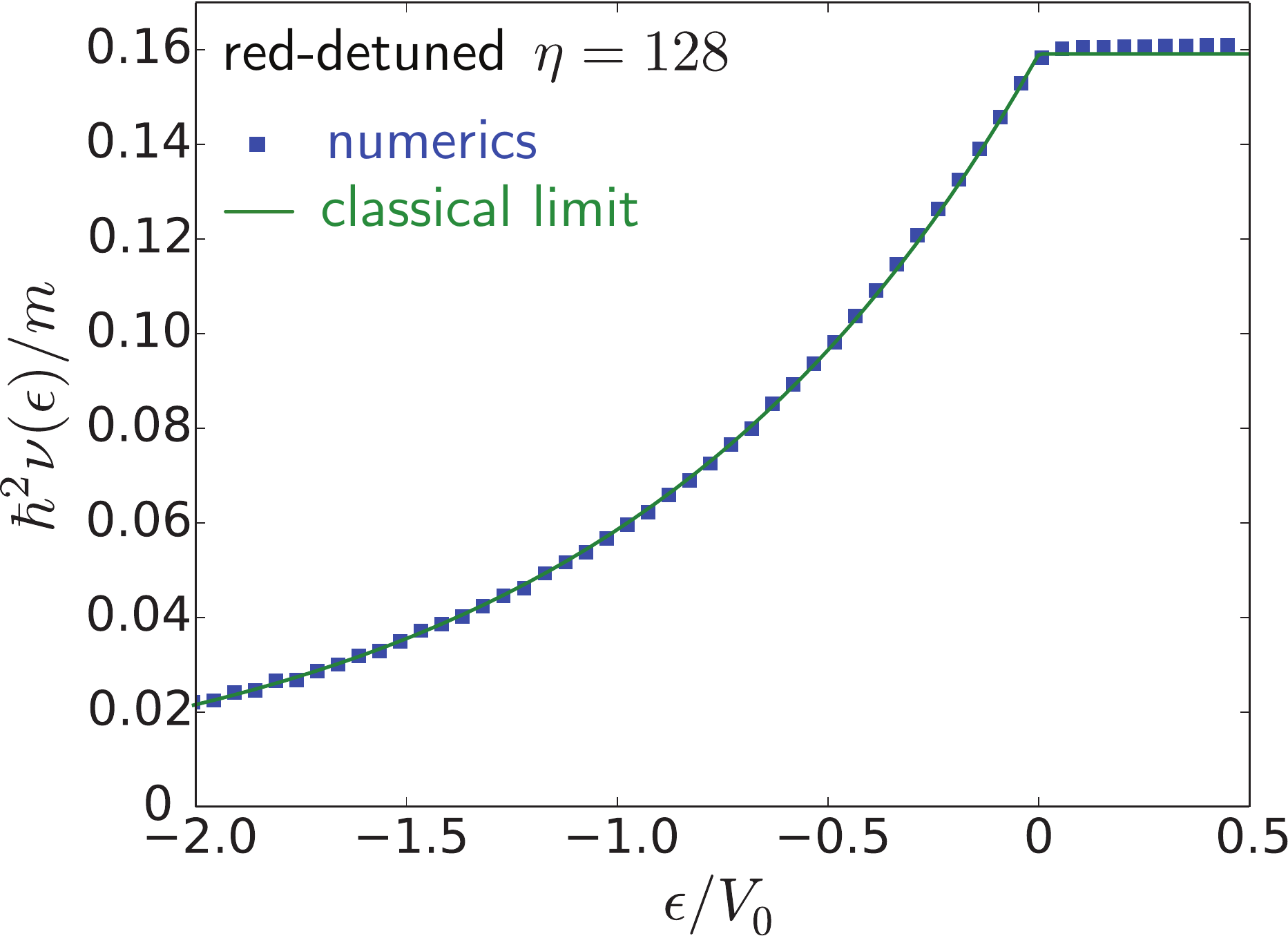}
\caption{(Color online) 
Density of states $\nu(\epsilon)$ as a function of energy in a 2D, red-detuned speckle potential with Gaussian correlation function. The classical limit, Eq. \eqref{eq-def-class-dos}, is shown as a solid green curve. Blue dots are the results from exact numerical simulations based on Eq. \eqref{eq-def-dos-x}.}
\label{fig-2Dreddos}
\end{figure}

We show in Fig. \ref{fig-2Dreddos} the DoS in a 2D, red-detuned speckle potential computed from numerical simulations using a system size $L \times L = (10 \pi \sigma)^2$ with $2000$ grid points in each direction and 8000 disorder realizations, based on Eq. \eqref{eq-def-dos-x} (blue dots).
As for 1D red-detuned speckles, the oscillator correction to the DoS diverges due to an ultraviolet divergence in the propagator, see Sec. \ref{sec-1D-red-DoS}.
Nevertheless, as seen in Fig. \ref{fig-2Dreddos}, the classical prediction \eqref{eq-def-class-dos} (solid green curve) already constitutes an excellent approximation of the exact result.

\section{Conclusion}
\label{secconcl}

In this paper, we have pointed out that an expansion in powers of $\hbar$ of the spectral function or the density of states in speckle potentials is not sufficient at low energies, due to the discontinuity of the potential distribution. In order to overcome this difficulty, we have developed a novel analytical method based on a semiclassical description of the dynamics
combined with the statistical properties of potential extrema. 
Applying this approach to 1D and 2D blue- and red-detuned speckles, we have carried out the calculation of the spectral function and the DoS.  
By connecting our results with those of previous works valid at high energies \cite{Trappe15, Falco10}, we have been able to describe the whole energy spectrum, and have found a good agreement with exact numerical simulations. 

Our semiclassical description additionally provides a simple interpretation of intriguing features of the spectral function and DoS. In particular, for blue-detuned potentials 
we have shown that the low-energy peak of spectral functions is essentially associated with the ground state of an atom in a potential well of the speckle, while the secondary bump is associated with excited states. 
We have also emphasized that in spite of their simple symmetry, red- and blue-detuned speckles exhibit remarkably different features in the semiclassical regime, coming from the fundamental different nature of the classical trajectories involved near zero energy: for blue-detuned speckles, these classical trajectories lie in deep potential wells, while for red-detuned speckles they lie in the vicinity of the top of inverted wells.  

As a logical continuation of this work, it would be of great interest to address the case of three-dimensional speckle potentials, involved in important questions related to Anderson localization \cite{Jendrzejewski12, Semeghini15, Pasek2015, Ghosh15}. This task appears challenging though, as the isolated points of zero potential in two dimensions become curves in three dimensions, making the application of a harmonic oscillator approximation less obvious.

\section*{Acknowledgments}

The authors would like to thank Thimoth\'ee Thiery and Cord M\"uller for useful discussions.
This work was granted access to the HPC
resources of TGCC under the allocation 2016-057083 made by GENCI (Grand
Equipement National de Calcul Intensif).

\section*{Appendix A} 
\label{appendixA}

In this appendix, we calculate the leading order smooth quantum corrections to the classical limit of the spectral function, Eq. \eqref{eq-semiclassics-Trappe}, using an alternative approach to the one used in \cite{Trappe15}. The calculation is first carried out for 1D, blue-detuned speckles, then generalized to any dimension, and finally to red-detuned speckles by a simple symmetry argument. 

The first stage of our approach is a commutator expansion of the evolution operator based on the Zassenhaus formula \cite{Magnus54}:
\begin{eqnarray}
&&\bra{k} e^{ - i\left[\frac{p^2}{2m} + V\right] t} \ket{k}
=   e^{ -iV(x)t } e^{ \frac{i\hbar^2 t^3}{3m} \left[ \partial_x V(x) \right]^2 }  \\
 && \times e^{   \frac{-\hbar^2 t^2}{4m} \left[ 2ik\partial_x V(x) + \partial_x^2 V(x)\right]
 -\frac{it^3 \hbar^2}{3m}   \left[ \partial_x^2 V(x) \right]\epsilon_k}  e^{O(\hbar^3)}.\nonumber
 \label{eq-commuexp}
\end{eqnarray}
The second stage consists in carrying out the disorder average. This can be done by mean of the following  cumulant expansion:
\begin{equation}
\overline{\text{exp} \left( X \right) } = \text{exp} \left[ \sum_{n=1}^{\infty} \frac{\kappa_n(X)}{n!} \right],
\end{equation}
where $\kappa_n$ denotes the $n^{\text{th}}$ cumulant. To evaluate the cumulants of sums of random variables that appear in Eq. \eqref{eq-commuexp}, we make use of the expansion
\begin{equation}
\kappa_n(X+Y) = \sum_{j=0}^n \dbinom{n}{j} \kappa(\underset{j \text{ terms}}{\underbrace{X, \dots,X}},\underset{n-j \text{ terms}}{\underbrace{Y, \dots,Y}}),
\end{equation}
where we have introduced the joint cumulants $\kappa$, defined as \cite{Brillinger01}
\begin{equation}
\kappa(X_1,\dots,X_n) = \sum_{\pi}( \vert \pi \vert -1)!(-1)^{\vert \pi \vert -1} \prod_{B \in \pi} \overline{ \left[  \prod_{i \in B} X_i \right] }.
\end{equation}
Here $\pi$ runs through the list of all partitions of $\{ 1, ..., n \}$, $B$ runs through the list of all blocks of the partition $\pi$, and $\vert \pi \vert$ is the number of parts in the partition. Joint cumulants have the following important properties \cite{Brillinger01}:
\begin{enumerate}
\item they are linear in all variables.

\item $\kappa(X, \dots, X)=\kappa_n(X)$.

\item $\kappa(X_1, \dots,X_n)=0$ if any set of the $X_i$'s are independent of the remaining $X_{j\ne i}$'s.
\end{enumerate}

After these premises, let us now write the random potential as
\begin{equation}
V(x) = E_1(x)^2 + E_2(x)^2,
\end{equation}
where $E_1$ and $E_2$ are independent Gaussian variables with zero mean and equal variance \cite{Goodman08}. Defining $X=-iV(x)t$ and denoting by $Y_i$ the $\hbar$ corrections appearing in Eq. \eqref{eq-commuexp}, we obtain for the $n^{\text{th}}$ cumulant:
\begin{widetext}
\begin{eqnarray}
\kappa_n\left( X+ \sum_{i=1}^{m} Y_i \right) & = & \kappa_n(X) + 
\sum_{i=1}^m n\kappa(\underset{n-1 \text{ terms}}{\underbrace{X,\dots,X}},Y_i) 
 \nonumber\\
& +& \sum_{j=1}^2 \dbinom{n}{2}  \kappa \left( \underset{n-2 \text{ terms}}{\underbrace{X,\dots,X}},-\frac{\hbar t^2}{2m}  \left[ \partial_x E_j^2(x) \right]  i\hbar k,-\frac{\hbar t^2}{2m}  \left[ \partial_x E_j^2(x) \right]  i\hbar k \right) + O \left( \hbar^3 \right).
\label{eq-cumulantn}
\end{eqnarray}
\end{widetext}
We now need to calculate the various cumulants entering this equation. 
For this purpose, we use a theorem due to Leonov and  Shiryaev \cite{Leonov59, footnote3}. Before discussing the theorem itself, it is useful to introduce some terminology. Consider the matrix
\\
\begin{equation}
\begin{pmatrix}
   X_{1 1} & \dots & X_{1 J} \\
   	. &  & . \\
   	. &  & . \\
   	. &  & . \\
   X_{J 1} & \dots & X_{J J}
   	\label{tableleonov}
\end{pmatrix},
\end{equation}
\\
and a partition $P_1 \cup P_2 \cup \dots \cup P_M$ of its entries. We choose this matrix square for simplicity, but the formalism is straightforwardly generalizable to rectangular matrices. If the rows are denoted by $R_1,\dots,R_J$, then a partition is said to be indecomposable if and only if there exist no sets $P_{m_1},\dots,P_{m_N}$, ($N<M$), and rows $R_{i_1},\dots,R_{i_P}$, ($P<J$), with
\begin{equation}
P_{m_1} \cup \dots \cup P_{m_N} = R_{i_1} \cup \dots \cup R_{i_P}.
\end{equation}
The theorem then goes as follows \cite{Leonov59}. Consider a matrix of random entries $X_{ij}$ ($i,j=1,\dots,J$) 
and the $J$ random variables
\begin{equation}
Y_i = \prod_{j=1}^J X_{ij}, \text{  } i=1,\dots,J.
\end{equation}
The joint cumulant $\kappa(Y_1,...,Y_J)$ is then given by
\begin{align}
\begin{split}
\kappa(Y_1,...,Y_J)= \sum_{P} & \kappa(\underset{\{i_1 j_1,\dots,i_m j_m\} = P_1}{\underbrace{X_{i_1 j_1},\dots,X_{i_m j_m}}}) \dots
\\ 
& \dots  \kappa(\underset{\{i_n j_n,\dots,i_o j_o\} = P_p}{\underbrace{X_{i_n j_n},\dots,X_{i_o j_o}}}),
\label{russianthm}
\end{split}
\end{align}
where the summation is over all indecomposable partitions $P = P_1 \cup \dots \cup P_p$ of matrix \eqref{tableleonov}. 

Let us now tackle one of the terms involved in Eq. \eqref{eq-cumulantn}:
 $\kappa(E_1^2,\dots,E_1^2,\partial_x^2 E_1^2)$. It is simpler to work in Fourier space, hence defining
\begin{equation}
E_j(x) = \int \frac{\text{d} p_i}{2 \pi}  e^{ip x} E_j(p).
\end{equation}
The cumulant of interest then reads
\begin{align}
\begin{split}
& \kappa(E_1^2,\dots,E_1^2,\partial_x^2 E_1^2)
 \\
& =  - \int \left[ \prod_{i=1}^{2n} \frac{ \text{d} p_i }{2 \pi } \right] (p_{2n-1} + p_{2n} )^2 
\kappa(E_1(p_1)E_1(p_2), \dots
\\
& \hspace{0.3cm} \dots, E_1(p_{2n-3})E_1(p_{2n-2}),E_1(p_{2n-1})E_1(p_{2n})).
\end{split}
\end{align}
The corresponding matrix \eqref{tableleonov} is
\begin{equation}
\begin{pmatrix}
   E_1(p_1) & E_1(p_2) \\
   . & .  \\
   . & .  \\
   . & .  \\
   E_1(p_{2n-1}) & E_1(p_{2n}) 
\end{pmatrix}.
\end{equation}
As $E_1$ is Gaussian distributed, only joint cumulants involving two fields should be kept in the right-hand side of Eq. \eqref{russianthm}. Our indecomposable partitions are then made of pairs of $E_1$ and all give the same contribution. Let us now count them, taking into account the two following constraints for making pairs so to obtain an indecomposable partition:
\begin{itemize}
\item a pair cannot be formed out of two fields lying on the same line, i.e. the choice
\begin{equation}
\begin{pmatrix}
 \circled{$E_1(p_1)$} & \circled{$E_1(p_2)$} \\
   . & .  \\
   . & .  \\
   . & .  \\
   E_1(p_{2n-1}) & E_1(p_{2n}) 
\end{pmatrix}
\end{equation}
is forbidden.
\item Two pairs right nearby cannot be formed, i.e. the choice
\begin{equation}
\begin{pmatrix}
   \circled{$E_1(p_1)$} & \squared{$E_1(p_2)$} \\
   \circled{$E_1(p_1)$} & \squared{$E_1(p_2)$} \\
   . & . \\
   . & . \\
   . & . \\
   E_1(p_{2n-1}) & E_1(p_{2n})
\end{pmatrix}
\end{equation}
is forbidden.
\end{itemize}
Therefore, to form the first pair, we have $(2n)$ choices for the first field and $(2n-2)$ choices for the second, and similarly for the next pairs. This leaves us with $2n(2n-2)^2(2n-4)^2 \dots =2^{2n} n!^2/(2n)$ choices of pairing. There is however a redundancy in this counting, due to the invariance of the partition with respect to both swapping of the two fields inside one pair ($2^n$ possibilities) and swapping of different pairs ($n!$ possibilities). This leaves us with only $2^n n!^2/(2n2^n n!)=2^n n!/(2n)$ choices of pairing. Calculating the contribution from one of them, we obtain
\begin{align}
\begin{split}
\kappa(\underset{n-1 \text{ terms}}{\underbrace{E_1^2, \dots,E_1^2}},\partial_x^2 E_1^2) = & \frac{2^n n!}{2n}   F^{n-2}(0) 2
\\
& \times
 \left[ F(0)F''(0) + F'^2(0) \right],
\end{split}
\end{align}
where $F(x)$ is defined by Eq. \eqref{Fx_def}. 

A similar derivation is then performed for all terms in Eq. \eqref{eq-cumulantn} that are not found to vanish on the basis of the property 2- above \cite{Brillinger01}. Upon summing various geometric series and recognizing the expansion of a logarithm, we find
\begin{align}
\begin{split}
&\overline{\bra{k} e^{ - i \left[ \frac{p^2}{2m} + V \right] \frac{t}{\hbar} } \ket{k} }
 = \frac{e^{-i\epsilon_k t /\hbar} }{1 \pm i tV_0/\hbar } \\
& \times \left[ 1 +  \frac{i t^3V_0^2E_\sigma/\hbar^3}{12 ( 1 + itV_0/\hbar)}  + \frac{t^4V_0^2E_\sigma/\hbar^4 }{12 (1 + itV_0/\hbar)}  \epsilon_k \right] ,
\end{split}
\end{align}
This result is not difficult to generalize to dimension $d$, where Eq. \eqref{eq-commuexp} becomes
\begin{eqnarray}
&&\bra{\bk} e^{ - i \left[ \frac{\bs{p}^2}{2m} + V \right] \frac{t}{\hbar} } \ket{\bk}
 =  e^{ -iV(\br)t }  e^{ \frac{i\hbar^2 t^3}{3m} \sum_{i=1}^d \left[ \partial_{x_i} V(\br) \right]^2 } \nonumber\\
 && \times e^{   \frac{-\hbar^2 t^2}{4m} \left[ 2i\bk\cdot  \boldsymbol{\nabla} V(\br) +  \boldsymbol{\nabla}^2 V(\br)\right] }\nonumber\\
 && \times e^{-\frac{i\hbar^2t^3 }{3m}  \sum_{i,j=1}^d \left[ \partial_{x_i} \partial_{x_j} V(\br) \right] \frac{ \hbar^2 k_i k_j }{2m} } e^{O(\hbar^3)}.
\end{eqnarray}
In the sum $\sum_{i,j=1}^d \left[ \partial_{x_i} \partial_{x_j} V(\br) \right]  \hbar^2 k_i k_j / 2m$, to leading order in  $\hbar$ the crossed terms ($i \neq j$) do not contribute to the disorder-averaged propagator as their contributions are proportional to first-order derivatives of the field correlation function \eqref{Fx_def} evaluated at $0$, which vanish. Also, derivatives of the potential with respect to different directions are independent. Therefore, the propagator in dimension $d$ is simply the product of $d$ 1D propagators. Finally, the result for the red-detuned speckle is deduced by changing $m$ to $-m$ and $t$ to $-t$ (which amounts to changing the sign of $V$). The general result then reads
\begin{eqnarray}
&&\overline{\bra{k} e^{ - i \left[ \frac{p^2}{2m} + V \right] \frac{t}{\hbar} } \ket{k} }
 =\frac{e^{-i\epsilon_k t /\hbar} }{1 \pm i tV_0/\hbar } \nonumber\\
&& \times \left[ 1 +  \frac{di t^3V_0^2E_\sigma/\hbar^3}{12 ( 1 \pm itV_0/\hbar)}  + \frac{t^4V_0^2E_\sigma/\hbar^4 }{12 (1 \pm itV_0/\hbar)}  \epsilon_k \right] ,
\end{eqnarray}
with the $+$ (resp. $-$) sign for blue-(resp. red-)detuned speckles. This immediately leads to Eq. \eqref{eq-semiclassics-Trappe} of the main text.

\section*{Appendix B}
\label{appendixB}

In this appendix, we derive the joint probability distribution $P(\lambda_1,\lambda_2)$ of the eigenvalues $\lambda_1$ and $\lambda_2$ of the matrix
\begin{equation}
A = 
\begin{pmatrix}
     \partial_x^2 V(x,y)  &  \partial_x \partial_y V(x,y)  \\
    \partial_y \partial_x V(x,y)   &  \partial_y^2 V(x,y)  \\
\end{pmatrix},
\end{equation}
in the vicinity of a minimum $V(x,y)=0$. As in Sec. \ref{Joint_dist_sec} we write the potential as
\begin{equation}
V(x,y) = \Re(x,y)^2 + \Im(x,y)^2,
\end{equation}
where $\Re(x,y)$ and $\Im(x,y)$ are independent Gaussian variables with zero mean and equal variance $\sigma_c^2=V_0/(4\sigma^2)$ \cite{Goodman08}. Making use of the short-hand notation 
\begin{align}
& \Re_x \equiv \partial_x \Re(x,y), & \Im_x \equiv \partial_x \Im(x,y), \nonumber\\
& \Re_y \equiv \partial_y \Re(x,y), & \Im_y \equiv \partial_y \Im(x,y), 
\end{align}
we rewrite the matrix $A$ as
\begin{equation}
A 
= 2
\begin{pmatrix}
     \Re_x^2 + \Im_x^2 &   \Re_x  \Re_y +  \Im_x \Im_y   \\
  \Re_x  \Re_y +  \Im_x \Im_y   & \Re_y^2 + \Im_y^2\\
\end{pmatrix}.
\label{matrixVxx}
\end{equation}
$\Re_x$, $ \Re_y$, $\Im_x$ and $\Im_y$ are independent, Gaussian distributed random variables with zero mean and variance $\sigma_c$.
The distribution $P(u,v)$ can then be expressed as
\begin{align}
\begin{split}
&P(u,v) = \int \text{d}\Re_x  \text{d}\Re_y \text{d}\Im_x \text{d}\Im_y P(\Re_x) P(\Re_y) P(\Im_x) P(\Im_y) \\
& \times \delta(u - \lambda_1(\Re_x,\Re_y,\Im_x,\Im_y) ) 
\delta(v - \lambda_2(\Re_x,\Re_y,\Im_x,\Im_y)).\nonumber
\end{split}
\end{align}
To tackle this integral, we first change variables to ``intensity'' and ``phase'':
\begin{align}
& \Re_x = \sqrt{I_1} \cos\theta_1, \hspace{0.5cm} \Im_x = \sqrt{I_1} \sin\theta_1,\nonumber \\
& \Re_y = \sqrt{I_2} \cos\theta_2, \hspace{0.5cm} \Im_y = \sqrt{I_2} \sin\theta_2.
\end{align}
The Jacobian of the transformation is $1/4$, and $I_1, I_2 \in \left[ 0, +\infty \right[$ and $\theta_1, \theta_2  \in \left[ -\pi, \pi \right]$. The integral reduces to
\begin{align}
\begin{split}
 &P(\lambda_1,\lambda_2) = \frac{1}{32 \pi^2 \sigma_c^4} \int_{0}^{+\infty} \text{d}I_1 \text{d}I_2 \int_{-\pi}^{\pi} \text{d}\theta_1 \text{d}\theta_2 e^{-\frac{I_1 + I_2}{2 \sigma_c^2} } 
\\
&  \times \delta \left( \lambda_1 - \left[ I_1 + I_2  - \sqrt{I_1^2+I_2^2 + 2I_1 I_2 \cos2 (\theta_1 - \theta_2)  }  \right]  \right)
\\
&  \times \delta \left( \lambda_2 - \left[ I_1 + I_2+ \sqrt{I_1^2+I_2^2 + 2I_1 I_2 \cos2 (\theta_1 - \theta_2)  } \right]  \right),\nonumber
\end{split}
\end{align}
where we have assumed $\lambda_2>\lambda_1$ without loss of generality, and added a corresponding renormalization prefactor $1/2$.
We then introduce
\begin{equation}
\varphi = \theta_1 + \theta_2, \hspace{0.5cm} \phi = 2(\theta_1 - \theta_2),
\end{equation}
and carry out the integral over $\varphi$. This eventually yields
\begin{align}
\begin{split}
& P(\lambda_1,\lambda_2) = \frac{1}{8 \pi \sigma_c^4} \int_{0}^{+\infty} \text{d}I_1 \text{d}I_2 \int_{0}^{\pi}  \text{d}\phi e^{-\frac{I_1 + I_2}{2 \sigma_c^2} } 
\\
& \hspace{0.3cm} \times \delta \left( \lambda_1 - \left[ I_1 + I_2 - \sqrt{I_1^2+I_2^2 + 2I_1 I_2 \cos\phi } \right] \right) 
\\
& \hspace{0.3cm} \times \delta \left( \lambda_2 - \left[ I_1 + I_2 + \sqrt{I_1^2+I_2^2 + 2I_1 I_2 \cos\phi  } \right] \right).\nonumber
\end{split}
\end{align}
This expression can be further simplified by writing $\int_0^\infty  \text{d}I_1 \text{d}I_2=\int_0^\infty  \text{d}I_1 \int_0^{I_1}\text{d}I_2+\int_0^\infty  \text{d}I_1 \int_{I_1}^\infty\text{d}I_2$ and noticing the equality of these two integrals due to the symmetric role played by $I_1$ and $I_2$:
\begin{align}
\begin{split}
& P(\lambda_1,\lambda_2) = \frac{1}{4 \pi \sigma_c^4} \int_{0}^{+\infty} \text{d}I_1 \int_{0}^{I_1}\text{d}I_2 \int_{0}^{\pi}  \text{d}\phi e^{-\frac{I_1 + I_2}{2 \sigma_c^2} } 
\\
& \hspace{0.3cm} \times \delta \left( \lambda_1 - \left[ I_1 + I_2 - \sqrt{I_1^2+I_2^2 + 2I_1 I_2 \cos\phi } \right] \right) 
\\
& \hspace{0.3cm} \times \delta \left( \lambda_2 - \left[ I_1 + I_2 + \sqrt{I_1^2+I_2^2 + 2I_1 I_2 \cos\phi  } \right] \right).\nonumber
\end{split}
\end{align}
We then change the variable $\phi$ to $z$ so that
\begin{equation}
z=I_1 + I_2 + \sqrt{I_1^2+I_2^2 + 2I_1 I_2 \cos\phi},
\end{equation}
where $z$ spans the interval $[0,2I_2]$. The corresponding Jacobian is
\begin{equation}
\abs{ \frac{\partial \phi}{\partial z} } = \frac{2\abs{I_1+I_2-z}}{\sqrt{z (2 I_1-z) (2 I_2-z)(2I_1+2I_2+z)}}.
\end{equation}
Performing the integrals over $I_2$ and $z$, we straightforwardly find
\begin{eqnarray}
P(\lambda_1,\lambda_2) &=&
\frac{1}{8 \pi \sigma_c^4} 
 \int_{\frac{\lambda_1+\lambda_2}{4}}^{\frac{\lambda_2}{2}}
  \text{d}I_1 e^{-\frac{\lambda_1 + \lambda_2}{4 \sigma_c^2} }  \nonumber\\
&&\times \frac{\left( \lambda_2 - \lambda_1 \right) \theta(\lambda_1)}{\sqrt{\lambda_1 \lambda_2 \left(  \lambda_2 - 2I_1 \right) \left( 2 I_1 - \lambda_1 \right)}}.
\end{eqnarray}
The remaining integral can be done analytically, yielding
\begin{equation}
\label{l2_sup_l1}
P(\lambda_1,\lambda_2) =
\frac{ \left( \lambda_2 - \lambda_1 \right) e^{-\frac{\lambda_1 + \lambda_2}{4 \sigma_c^2} }  }{32 \sigma_c^4\sqrt{\lambda_1 \lambda_2 }}
\theta(\lambda_1)\ \ (\lambda_2>\lambda_1).
\end{equation}
This relation has been obtained assuming $\lambda_2>\lambda_1$. The opposite case $\lambda_1<\lambda_2$ is fully symmetric:
\begin{equation}
\label{l1_sup_l2}
P(\lambda_1,\lambda_2) =
\frac{ \left( \lambda_1 - \lambda_2 \right) e^{-\frac{\lambda_1 + \lambda_2}{4 \sigma_c^2} }  }{32 \sigma_c^4\sqrt{\lambda_1 \lambda_2 }}
\theta(\lambda_2) \ \ (\lambda_2<\lambda_1).
\end{equation}
Using Eqs. \eqref{l2_sup_l1} and \eqref{l1_sup_l2} together with the relations $\lambda_1=m\omega_x^2$, $\lambda_2=m\omega_x^2$, we finally obtain Eq. \eqref{P2D_cond} of the main text.


\begin{thebibliography}{99}
\bibliographystyle{unsrt}

\bibitem{Anderson1958}
P.W. Anderson, \href{http://journals.aps.org/pr/abstract/10.1103/PhysRev.109.1492}{Phys. Rev. \textbf{109}, 1492 (1958)}.

\bibitem{Chabe08}
J. Chab\'e, G. Lemari\'e, B. Gr\'emaud, D. Delande, P. Szriftgiser and J. C. Garreau, 
\href{http://journals.aps.org/prl/abstract/10.1103/PhysRevLett.101.255702}{Phys. Rev. Lett. \textbf{101}, 255702 (2008)}. 

\bibitem{Manai15}
I. Manai, J. F. Cl\'ement, R. Chicireanu, C. Hainaut, J. C. Garreau, P. Szriftgiser and D. Delande,
\href{http://journals.aps.org/prl/abstract/10.1103/PhysRevLett.115.240603}{Phys. Rev. Lett. \textbf{115}, 240603 (2015).}

\bibitem{Billy2008}
J. Billy, V. Josse, Z. Zuo, A. Bernard, B. Hambrecht, P. Lugan, D. Cl\'ement, L. Sanchez-Palencia, P. Bouyer and A. Aspect, \href{http://www.nature.com/nature/journal/v453/n7197/abs/nature07000.html}{Nature \textbf{453}, 891 (2008)}.

\bibitem{Jendrzejewski12}
F. Jendrzejewski, A. Bernard, K. M\"uller, P. Cheinet, V. Josse, M. Piraud, L. Pezz\'e, L. Sanchez-Palencia, A. Aspect and P. Bouyer, 
\href{http://www.nature.com/nphys/journal/v8/n5/full/nphys2256.html}{Nature Phys. \textbf{8}, 398 (2012)}.

\bibitem{Semeghini15}
G. Semeghini, M. Landini, P. Castilho, S. Roy, G. Spagnolli, A. Trenkwalder, M. Fattori, M. Inguscio and G. Modugno, \href{http://www.nature.com/nphys/journal/v11/n7/abs/nphys3339.html}{Nature Phys. \textbf{11}, 554 (2015)}.

\bibitem{Hu2008}
H. Hu, A. Strybulevych, J. H. Page, S. E. Skipetrov and  B. A. Van Tiggelen, \href{http://www.nature.com/nphys/journal/v4/n12/abs/nphys1101.html}{Nature Phys. \textbf{4}, 945 (2008)}.

\bibitem{Modugno10}
G. Modugno, 
\href{http://iopscience.iop.org/article/10.1088/0034-4885/73/10/102401/meta;jsessionid=0F969B00887371C539D956F431B575B7.c3.iopscience.cld.iop.org}{Rep. Prog. Phys. \textbf{73}, 102401 (2010)}.

\bibitem{Shapiro12}
B. Shapiro,
\href{http://iopscience.iop.org/article/10.1088/1751-8113/45/14/143001}{J. Phys. A: Math. Theor. \textbf{45}, 143001 (2012)}.

\bibitem{Anderson97}
P. W. Anderson, Basic Notions of Condensed Matter Physics (Westview Press / Addison-Wesley, 1997).

\bibitem{McGehee13}
W. R. McGehee, S. S. Kondov, W. Xu, J. J. Zirbel, and B. DeMarco,
\href{http://journals.aps.org/prl/abstract/10.1103/PhysRevLett.111.145303}{Phys. Rev. Lett. \textbf{111}, 145303 (2013)}.

\bibitem{Mueller14}
C. A. M\"uller and B. Shapiro,
\href{http://journals.aps.org/prl/abstract/10.1103/PhysRevLett.113.099601}{Phys. Rev. Lett. \textbf{113}, 099601 (2014).}

\bibitem{Mueller16}
C. A. M\"uller, D. Delande and B. Shapiro, submitted to Phys. Rev. A (2016), \href{https://arxiv.org/abs/1605.04329}{arxiv:1605.04329.} 

\bibitem{Lugan07}
P. Lugan, D. Clement, P. Bouyer, A. Aspect and L. Sanchez-Palencia, 
\href{http://journals.aps.org/prl/abstract/10.1103/PhysRevLett.99.180402}{Phys. Rev. Lett. \textbf{99},180402 (2007)}.

\bibitem{Aleiner10}
I. L. Aleiner, B. L. Altshuler, and G. V. Shlyapnikov,
\href{http://www.nature.com/nphys/journal/v6/n11/abs/nphys1758.html}{Nature Phys. \textbf{6}, 900 (2010)}.

\bibitem{Cherroret15}
N. Cherroret, T. Karpiuk, B. Gr\'emaud and C. Miniatura, 
\href{http://journals.aps.org/pra/abstract/10.1103/PhysRevA.92.063614}{Phys. Rev. A \textbf{92}, 063614 (2015).}

\bibitem{Trappe15}
M. I. Trappe, D. Delande and C. A. M\"uller,
\href{http://iopscience.iop.org/article/10.1088/1751-8113/48/24/245102}{J. Phys. A: Math. Theor. \textbf{48}, 245102 (2015)}.

\bibitem{Falco10}
G. M. Falco, A. A. Fedorenko, J. Giacomelli and M. Modugno, 
\href{http://journals.aps.org/pra/abstract/10.1103/PhysRevA.82.053405}{Phys. Rev. A \textbf{82}, 053405 (2010)}.

\bibitem{John88}
S. John, M. Y. Chou, M. H. Cohen and C. M. Soukoulis, 
\href{http://journals.aps.org/prb/abstract/10.1103/PhysRevB.37.6963}{Phys. Rev. B \textbf{37}, 6963 (1988)}.

\bibitem{Goodman08}
J. W. Goodman, Statistical Properties of Laser Speckle Patterns (Springer, 2008).

\bibitem{Mathematica}
Wolfram Research, Inc., Mathematica, Version 10.3, Champaign, IL (2015).

\bibitem{Kuhn07}
R. C. Kuhn, O. Sigwarth, C. Miniatura, D. Delande, and C. A. M\"uller,
\href{http://iopscience.iop.org/article/10.1088/1367-2630/9/6/161}{New J. Phys. \textbf{9}, 161 (2007)}.

\bibitem{Grammaticos79}
B. Grammaticos and A. Voros, \href{http://www.sciencedirect.com/science/article/pii/0003491679903439}{Ann. Phys. \textbf{123}, 359 (1979)}.

\bibitem{Gutzwiller90}
M. C. Gutzwiller, Chaos in classical and Quantum mechanics (Springer, 1990).

\bibitem{Tabor83}
M. Tabor,
\href{http://www.sciencedirect.com/science/article/pii/0167278983900052}{Physica \textbf{6D}, 195 (1983)}.

\bibitem{Haake10}
F. Haake, Quantum Signatures of Chaos (Springer, 2010).

\bibitem{footnote1}
Note that for blue-detuned speckle potentials, the spectral function and the density of states exhibit a Lifshitz tail  at very small energies \cite{Lifshitz64, Falco15}. This tail is intrinsically quantum and cannot be described by semiclassical approximations. 

\bibitem{Lifshitz64}
I. M. Lifshitz, \href{http://www.tandfonline.com/doi/abs/10.1080/00018736400101061}{Adv. Phys. \textbf{13}, 483 (1964)}.

\bibitem{Falco15}
G. M. Falco and A. A. Fedorenko,
\href{http://journals.aps.org/pra/abstract/10.1103/PhysRevA.92.023412}{Phys. Rev. A \textbf{92}, 023412 (2015)}.

\bibitem{Altland10}
A. Altland and B. Simons, Condensed Matter Field Theory (Cambridge University Press, 2010).

\bibitem{footnote1b}
It could be possible to perform a Wick rotation and express everything as a sum of imaginary ``energies''. This treatment is not very enlightning.

\bibitem{footnote0}
The maxima of blue-detuned speckle -- or equivalently the minima of a red-detuned one -- could be studied along the same lines, but they are not relevant for the behavior of the spectral function.

\bibitem{Watson66}
G. N. Watson, A treatise on the theory of Bessel functions (Cambridge University Press, 1966).

\bibitem{Weinrib82}
A. Weinrib, \href{http://journals.aps.org/prb/abstract/10.1103/PhysRevB.26.1352}{Phys. Rev. B \textbf{26}, 1352 (1982)}.

\bibitem{Halperin82}
A. Weinrib and B. I. Halperin, 
\href{http://journals.aps.org/prb/abstract/10.1103/PhysRevB.26.1362}{Phys. Rev. B \textbf{26}, 1362 (1982)}.

\bibitem{Gutzwiller82}
M. C. Gutzwiller, \href{http://www.sciencedirect.com/science/article/pii/0167278982900173}{Physica D \textbf{5}, 183 (1982).}

\bibitem{footnote2}
We remind that this value corresponds to a speckle potential with Gaussian correlation function. For circular speckles \cite{Goodman08}, this number goes up to roughly $75 \%$.

\bibitem{Pasek2015}
M. Pasek, Z. Zhao, D. Delande, G. Orso, \href{http://journals.aps.org/pra/abstract/10.1103/PhysRevA.92.053618}{Phys. Rev. A \textbf{92}, 053618 (2015).}

\bibitem{Ghosh15}
S. Ghosh, D. Delande, C. Miniatura, and N. Cherroret, 
\href{http://journals.aps.org/prl/abstract/10.1103/PhysRevLett.115.200602}{Phys. Rev. Lett. \textbf{115}, 200602 (2015).}

\bibitem{Magnus54}
W. Magnus, \href{http://onlinelibrary.wiley.com/doi/10.1002/cpa.3160070404/abstract;jsessionid=07A712665B526938D3783C4B48928CB8.f04t03}{Comm. Pure Appl. Math. \textbf{7}, 649 (1954)}.

\bibitem{Brillinger01}
D. R. Brillinger, Time series data analysis and theory (Siam, 2001).

\bibitem{Leonov59}
V. P. Leonov and A. N. Shiryaev,
\href{http://epubs.siam.org/doi/abs/10.1137/1104031}{Theo. prob. and appl. \textbf{4}, 319 (1959)}. Note that ``semi-invariant'' is another word for joint cumulant. 

\bibitem{footnote3}
Here we use a somewhat simpler and less general version of Leonov and Shiryaev theorem, proposed in \cite{Brillinger01}.


\end{thebibliography}
\end{document}